\documentclass[a4paper,useAMS,usenatbib,fleqn]{mnras}
\usepackage{graphics,graphicx}
\usepackage{amsmath}
\usepackage{hyperref}
\usepackage{amssymb}
\usepackage{natbib}
\usepackage{aas_macros}
\usepackage[usenames,dvipsnames]{xcolor}
\newcommand{\eagle}{{\sc eagle}}

\renewcommand{\aap}{A\&A}

\title[CEMPlifying reionization]{CEMPlifying reionization}

\author[]
{Mahavir Sharma\thanks{mahavir.sharma@durham.ac.uk}, Tom Theuns \& Carlos Frenk\\
Institute for Computational Cosmology,  Department of Physics, Durham University, South Road, Durham, UK, DH1 3LE \\
}

\begin{document}

\date{Submitted ---------- ; Accepted ----------; In original form ----------}

%\pagerange{\pageref{firstpage}--\pageref{lastpage}} 

\maketitle

\begin{abstract}
The massive stars that ionised the Universe have short lifetimes and can only be studied near the time of formation, but any low mass stars that formed contemporaneously might be observable in the local Universe today. We study the abundance pattern and spatial distribution  of these \lq siblings of reionizers\rq\ (SoRs) in the \eagle\ cosmological hydrodynamical simulation. SoRs  tend to be enriched to super-solar levels in $\alpha$-elements compared to iron.  In {\sc eagle} galaxies resembling the Milky Way, $\sim 40$~percent of carbon-enhanced metal poor (CEMP) stars are SoRs. Conversely,  $\sim 10$~percent of all SoRs are CEMP stars. This fraction increases to $\gtrsim 50$~percent  for SoRs of metallicity [Fe/H]$<-4$, and at such low metallicities, most of the CEMP stars are of CEMP-no subtype that are lacking neutron capture elements. Although these numbers may well depend on the details of the physical models implemented in {\sc eagle}, the trends we describe are robust as they result from the strong feedback from star formation in early galaxies, itself a key ingredient of most current models of galaxy formation. We further find that most SoRs today reside in  halos with mass $M_h\gtrapprox 10^{12}$~M$_\odot$, and 50~percent of them are in the halo of their central galaxy (distance $>10$~kpc), mainly because they were accreted onto their current host rather than formed in-situ. To a good approximation, the SoRs are CEMP-no stars that reside in the stellar halos of massive galaxies, with nearly half of them contributing to the intracluster light in groups and clusters.
\end{abstract}
%\hspace{0.1mm}
\section{Introduction}
The measurement of the Thomson optical depth towards the surface of last scattering suggests that the intergalactic medium was completely reionized between redshift 6 and 9 \citep{Planck15}. The details of this process are uncertain and the nature of the sources that dominated the production of ionising photons is hotly debated, with \lq first\rq\ stars (with zero-metallicity and thought to be relatively massive) and the first generation of galaxies, the leading candidates \citep[e.g.][]{Bouwens15, Robertson15,Sharma16}. The contribution of quasars is controversial (compare e.g. \citealt{Haardt15} to \citealt{Parsa17}). Studying these candidate reionizers in detail is one of the main science drivers	of the {\em James Webb Space Telescope} \citep[JWST,][]{Gardner06}; see {\em e.g.} \cite{Zackrisson17}.

An alternative way of learning about the source of reionization is through \lq galactic archaeology\rq (or \lq near field cosmology\rq): studying the properties of the oldest stars, in particular those that formed before $z\gtrapprox 6$, in the local Universe \citep[e.g.][]{Trenti10,Frebel15}. High-mass stars, the likely dominant sources of ionising photons, have short lifetimes ($\sim 20$~Myr), and hence only their remnants - black holes or neutron stars if they have remnants at all - would exist today; therefore the \lq reionizers\rq\ themselves cannot be studied locally. However, lower-mass stars that formed contemporaneously with the reionizers may survive until the present day, and be detectable in the Milky Way and its dwarf satellites. We will refer to such stars as the \lq siblings of reionizers\rq\ (SoRs for short). Their numbers, spatial distribution, and composition, may provide valuable clues to star formation during the epoch of reionization. Determining the ages of such old stars accurately is challenging, but their composition - the abundance of elements heavier than helium\footnote{We will use the symbol $Z$ to refer to the mass fraction of such elements.} - can be used as a proxy for age and it can provide a wealth of information on the assembly history of galaxies and the epoch of reionization \citep[e.g.][]{Beers05,Bromm11,Beers05,Frebel15}. Specific stellar populations (e.g. SoRs) can then be identified by searching for the expected abundance patterns in metal poor stars \citep[e.g.][]{Frebel08}.

A number of surveys have searched the Milky Way and its dwarf satellites for metal poor stars \citep[see][for a review]{Frebel15}. \cite{Bond70} conducted one of the first surveys to identify metal poor stars in the Galactic halo. \cite{Beers85} (see also \citealt{Beers92}) undertook an objective prism survey to identify low metallicity stars. 
More recently, the ESO-Hamburg Survey \citep{Christlieb03} and SEGUE/SDSS survey \citep{Caffau11b} detected a large number of stars with [Fe/H]$<-3$ \citep{Norris13a} and a few with even lower abundance, up to [Fe/H]$<-5$ \citep{Christlieb02,Frebel08,Keller14}, which interestingly appear to be the oldest stars discovered to date \citep{Frebel15}. Many more stars will be discovered by upcoming surveys such as LAMOST \citep{Li15} and GAIA \citep{Gaia16}. The peculiar metallicity pattern of some of the detected stars has been suggested to point to enrichment by very early low-$Z$ supernova \citep[e.g.][]{Aoki14}.

Rather than studying individual peculiar stars, it might be possible to study the abundances of large numbers of old stars and constrain the nature of reionizers as well as the properties of their host galaxy. \cite{Sharma16b} used simulations to demonstrate that the bursty nature of star formation in galaxies at redshifts $z\gtrapprox 6$, combined with poor metal mixing in early enrichment events, gives rise to characteristic abundance patterns at low metallicity. In these simulations, \cite{Sharma16b} found that the first generation of SoRs is enriched by massive low metallicity type II supernovae (SNe), whose yields exhibit large over abundances of carbon compared to iron (usually referred to as \lq CEMP\rq\ stars - carbon enhanced metal poor stars) amongst other abundance peculiarities. Strong outflows powered by these same SNe then temporarily suppress further star formation in the young galaxy and enrich the surroundings with type~II ejecta. Such winds are thought to be the origin of the metals detected in the intergalactic medium \cite[e.g.][]{Becker06, Cai17}.

Star formation can resume once the galaxy accretes gas as part of the galaxy's build-up. The newly accreted pristine gas can be enriched with carbon produced by asymptotic giant branch stars (AGB stars), yielding a second population of SoRs that are carbon enhanced compared to iron and to oxygen as well, additionally exhibiting s-process characteristics. The episodic inflow and outflows of gas in galaxies during their first star formation episodes - which we refer to as breathing modes - combined with the yields of stars at low metallicity, therefore predicts a large spread in [C/O] at low $Z$, with stars with high [C/O] that were enriched by AGB stars displaying additionally evidence for s-process burning, and stars with lower [C/O] displaying evidence for low-$Z$ type~II nucleosynthesic processing, such as CEMP-no characteristics (see \citealt{Beers05} for an overview of the properties and nomenclature of CEMP stars, and, in particular, the definition of CEMP-s (for s-process enriched), CEMP-r (for r-process enriched), and CEMP-no stars). \cite{Sharma16b} argue that the abundance pattern measured in low-$Z$ Milky Way stars provides some support for this sequence of events. In this paper, we further investigate the connection between SoRs and CEMPs, by studying their chemical abundance patterns, statistics and spatial distributions in simulations.

The theoretical studies by \citet{White00} and \citet{Brook07} have used dark matter only cosmological simulations to track dark matter particles judiciously chosen to be SoRs. This technique has also been adopted in semi-analytic models \citep[e.g.][]{Salvadori10,Tumlinson10}. These studies suggest that the oldest Milky Way stars reside in the bulge - where they might be difficult to detect given the high stellar density there. Hydrodynamical simulations show that fragments of pre-reionization galaxies containing SoRs may be contained in local group dwarf galaxies today (\citealt{Gnedin06}; see also \citealt{Madau08,Bovill11}).

Simulations that include gas physics such as e.g. \eagle\ \citep{Schaye15}, do not have resolution at the scale of a star, but still can identify the fossils of early galaxies in the present day Universe and hence provide a clue to where to look for SoRs. \cite{Starkenburg17} used the {\sc apostle} simulation \citep{Sawala16} and found that though the overall population of metal poor old stars is centrally concentrated, the fraction of such stars to all stars, increases with distance from the centre. They further find that more than half of the most metal poor stars should be located outside the solar circle. The main driver of these trends has not been examined in detail yet, with radial migration of stars and the manner in which the MW accretes SoRs, both likely playing a role. These mechanisms were invoked by {\cite{Navarro17}  (but see \citet{Haywood15}) to explain the separation between the thin and the thick discs in the [$\alpha$/Fe] versus [Fe/H] diagrams for MW stars \citep{Bovy16}.

In this paper, we use the \eagle\ cosmological hydrodynamical simulation \citep{Schaye15,Crain15} to identify SoRs among the metal poor stars. We further examine the spatial distribution of SoRs in galaxies and examine the processes that give rise to the trends that we find. The paper is structured as follows. Section 2 starts with a brief overview of the \eagle\ simulations, and continues with a discussion of the key aspects of our model for reionizers and their siblings. In Section 3, we investigate the connection between SoRs and the observed stars in the Milky way. In Section 4, we explore the distribution of stars in other galaxies. We summarise our findings in Section 5.

\section{Simulations and model assumptions}
\subsection{The \eagle\ simulation}
\label{sec_sim}
\eagle\ \citep{Schaye15} is a suite of cosmological hydrodynamical simulations based on the {\sc gadget} smoothed particle hydrodynamic (SPH) code \citep{Springel05}. The simulation code includes modification to the hydro-solver to resolve known issues with standard SPH, as well as a set of \lq subgrid\rq\ physics modules to capture the unresolved physics of interstellar medium, and of feedback from stars and accreting supermassive black holes. Numerical parameters that appear in these modules are calibrated to $z\approx 0$ observations of galaxies, in particular the galaxy stellar mass function, the relation between galaxy mass and galaxy size, and the relation between galaxy mass and black hole mass, see \cite{Crain15} for details. Galaxies are identified in post-processing with the {\sc subfind} algorithm described by \cite{Springel01, Dolag09}.

We briefly discuss the subgrid modules here (see \citealt{Schaye15} for full details). 
A spatially uniform but evolving background of UV, X-ray, CMB and ionising photons, as computed by \cite{Haardt01}, is switched on at redshift 11 to mimic reionisation of hydrogen and of \ion{He}{I} to \ion{He}{II}. The ionising radiation photo-heats gas and suppresses radiative cooling, the implementation of this process in \eagle\ is discussed in detail by \cite{Wiersma09a}. The implementation of star formation follows \cite{Schaye08}: a collisional gas particle is stochastically converted to a collisionless star particle above a metallicity-dependent density threshold, at a rate that guarantees that simulated galaxies follow the Kennicutt-Schmidt law \citep{Kennicutt98}. Star particles in the simulation represent a single population of stars with a \cite{Chabrier03} initial mass function (IMF) in the mass range $0.1\hbox{--}100$~M$_\odot$. Feedback from star formation is implemented using a scheme in which the loss of energy due to lack of resolution is compensated for by the additional injection of energy, see \citealt{Crain15}. Stellar evolution is implemented as described by \cite{Wiersma09b}. The simulation traces the abundances of 11 elements (H, He, C, N, O, Ne, Mg, Si, S, Ca, Fe) and tracks how these are burned and synthesised by AGB stars, type Ia SNe, and massive stars and their type II SNe end stages, see \cite{Wiersma09b} for a description of the stellar life-times and metallicity dependent yields used. Many properties of the simulated galaxies can be queried in the public data base described by \cite{McAlpine16}, and the particle data of the simulation has been made public as well \citep{Eagle17}.
	
In this paper we use the \eagle\ simulation labelled Recal-L025N0752  in table~2 of \cite{Schaye15}. The cubic simulation volume has comoving sides of 25~Mpc, and the simulation has dark matter particles of mass $1.21\times10^6$~M$_\odot$ and gas particles with initial mass $2.26\times10^5$~M$_\odot$; the Plummer-equivalent gravitational softening is $\epsilon=0.35~{\rm kpc}$ at $z=0$. We refer to this particular member of the \eagle\ simulation suite as \lq the \eagle\ simulation below.

\subsection{Reionizers in \eagle, and their siblings}
\label{sec_def}
Our model for reionization is described by \citet{Sharma16,Sharma17}, we begin with a brief summary of this work. The model postulates that the escape fraction of ionising photons increases with redshift. Such a rapid increase is required to simultaneously explain the low values of $f_{\rm esc}$ measured in the local Universe, not overproduce the amplitude of the ionising background at redshifts $z=1-4$, and yet have galaxies emit the required number of ionising photons to reionise the Universe by $z\gtrsim 6$ \citep[e.g.][]{Haardt12, Khaire15}. In our model, the increase in $f_{\rm esc}$ is a consequence of star formation (in {\sc eagle} and many other simulations, and presumably also in the Universe) becoming increasingly bursty at high $z$, with bursts creating channels in the galaxy's interstellar medium through which ionising photons can escape. The model follows \cite{Heckman01} in linking such bursty outflows to regions where star formation occurs where the surface density\footnote{$\dot\Sigma_\star$ is the surface density of star formation averaged on $~\sim {\rm kpc}^2$,  scale, similar to the star formation surface density appearing in the Kennicutt-Schmidt law  \citep{Kennicutt98}.} of star formation, $\dot\Sigma_\star$, is higher than a threshold value of $\dot \Sigma_{\rm \star,crit}\approx 0.1 {\rm M}_\odot {\rm yr}^{-1}~{\rm kpc}^{−2}$. The model further assumes that when $\dot\Sigma_\star<\dot \Sigma_{\rm \star,crit}$, the escape fraction of ionising photons is low, $f_{\rm esc}\approx 0$, whereas when $\dot\Sigma_\star\geq \dot \Sigma_{\rm \star,crit}$, $f_{\rm esc}=20$~per cent. 

There is some observational evidence to support the basic assumptions of our model. Most star formation in the local Universe occurs at surface densities $\dot\Sigma_\star<< \dot \Sigma_{\rm \star,crit}$ \citep{Brisbin12}, hence the model predicts $f_{\rm esc}\approx 0$, consistent with the low values inferred observationally which are of order a few per cent or less \citep[e.g.][]{Bland-Hawthorn01, Bridge10, Mostardi15}, even up to $z\sim 1$. In contrast high values of $f_{\rm esc}\sim 20$~per cent are measured in the few local cases of extremely compact vigorously star forming galaxies \citep{Borthakur14, Izotov16, Leitherer16, Marchi17}, which indeed have $\dot\Sigma_\star\gg \dot \Sigma_{\rm \star,crit}$. There is growing evidence that such compact galaxies with high star formation rate, and hence high values of $\dot\Sigma_\star$, indeed are associated with strong outflows or at least have a strongly turbulent interstellar medium \citep[e.g.][]{Amorin17, Chrisholm17}, consistent with such turbulent motions creating channels allowing ionising photons to escape. The bursty nature of star formation in small galaxies at high-$z$ is well established in high-resolution numerical simulations \cite[e.g.][]{Wise09, Kimm14, ElBadry16} where it allows ionising photons to escape \citep{Trebitsch17}. In addition, the detection of metals in the intergalactic medium at $z\gtrapprox 6$ \citep[e.g.][]{Becker06} provides fossil evidence 	that outflows are indeed ubiquitous at high $z$.

%Even though the simulations by \cite{Trebitsch17} support  the view that bursty feedback from star formation creates channels for the escape of ionising photons,	there is little consensus on the value of $f_{\rm esc}$ within the simulation community. Contrast for example the varied predictions for $f_{\rm esc}$ from the simulations by \cite{Kimm14} ($f_{\rm esc}\sim 20$~per cent), \cite{Ma15} ($f_{\rm esc}=5$~per cent), and \cite{Paardekooper15} (even lower $f_{\rm esc}$). This lack of guidance motivates us to set $f_{\rm esc}$ to 20~per cent in bursty galaxies, based on the observed value quoted by \cite{Borthakur14} and \citet{Izotov16}. Finally, although massive stars may have cleared channels to allow photons to escape, it may be that lower-mass stars in binaries may be the dominant source of the ionising photons themselves \citep{Stanway16}, relieving the tension between the time needed to clear channels and the short life-times of massive stars; see also \citet{Ma16}.
%
	
Galaxies at high redshift are small and compact in \eagle\ \citep{Furlong17,Sharma17} as well as in the observed Universe \citep[e.g.][]{Shibuya15}. We find that more than 80~percent of star formation at $z\gtrsim 6$ in \eagle\ occurs at $\dot\Sigma_\star> \dot \Sigma_{\rm \star,crit}$, which we take to yield $f_{\rm esc}\sim 20$~percent. Given this we will assume in this paper that all star formation before $z=6$ contributes to reionization. We then divide stars formed at $z\gtrapprox 6$ into two categories, \lq reionizers\rq\ - those stars that produced the ionising photons and which have no age  datable remnants today - and their siblings (SoRs) - stars that formed contemporaneously and are of low enough mass to survive to the present. Although SoRs do not necessarily have extremely low metallicity, their metallicity patterns may nevertheless provide evidence for enrichment by reionizers. Since detailed observations of individual stars is only possible locally, we begin by examining SoRs in Milky Way-like \eagle\ galaxies.

\section{S\lowercase{o}R\lowercase{s}  in the Milky Way}
\begin{figure}
 \centering
 \includegraphics[width=\linewidth]{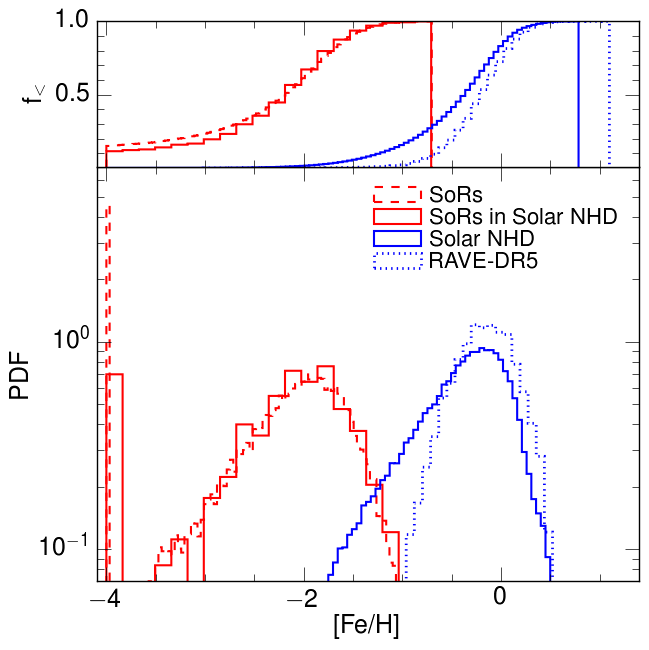}
 \caption{{\em Bottom panel}: Distribution of [Fe/H] for stars in galaxies that inhabit a halo of mass $10^{12}< M_{h}<10^{12.5}$~$h^{-1}$M$_\odot$ and with star formation rate $\dot M_\star \gtrsim 1$~M$_\odot$~yr$^{-1}$ (\lq Milky Way-like\rq\ galaxies) at $z=0$ in the RecalL0025N0752 \eagle\ simulation. The {\em blue} line refers to the stars in the solar neighbourhood ($6<R<10$~kpc, $|z|<2$~kpc) in such galaxies, the {\em red solid line} refers to such stars that additionally formed before $z=6$ (termed siblings of reionizers, or SoRs); the red dashed line to all SoRs in MW type galaxies. The {\em blue dotted line} shows the observed distribution from the {\sc rave} data release 5 \citep{Kunder17}. {\em Top panel}: corresponding cumulative distributions. In \eagle, approximately 60~percent of SoRs are metal poor ([Fe/H]$<-2$), whereas less than 1~per cent of all stars are metal poor. The apparent over-abundance of stars with [Fe/H]$<-1$ in \eagle\ simulation compared to that in {\sc rave} may be due to a lack of metal mixing in the simulations, see text for discussion and \citet{Kunder17} for details of observational biases.}
 \label{fig_FeH}
\end{figure}
\begin{figure}
 \centering
 \includegraphics[width=\linewidth]{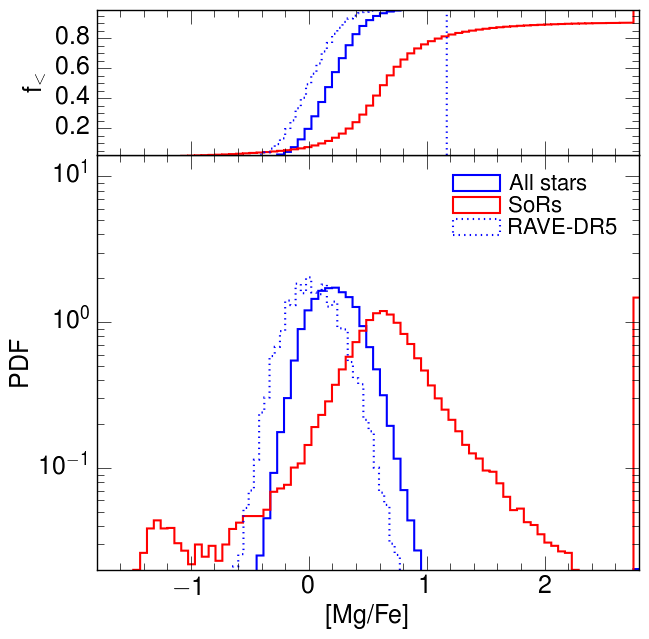}
\caption{{\em Bottom panel}: Distribution of [Mg/Fe] in \eagle\ MW stars.
	The {\em blue} line refers to all the stars, the {\em red solid line} refers to stars that  formed before $z=6$. The distribution of [Mg/Fe] of {\sc rave} stars \citep{Kunder17} is shown as a {\em blue dotted line}. {\em Top panel:} corresponding cumulative distributions. The distribution of [Mg/Fe] in \eagle\ is very similar in shape to that measured by {\sc rave}, but offset by $\sim 0.1$~dex to higher values. SoRs have a much wider distribution in [Mg/Fe], are typically enhanced in [Mg/Fe], and display a striking {\em minimum} value of [Mg/Fe]$\gtrapprox -1.5$ (see text for discussion). }
 \label{fig_MgFe}
\end{figure}
\begin{figure}
 \centering
 \includegraphics[width=\linewidth]{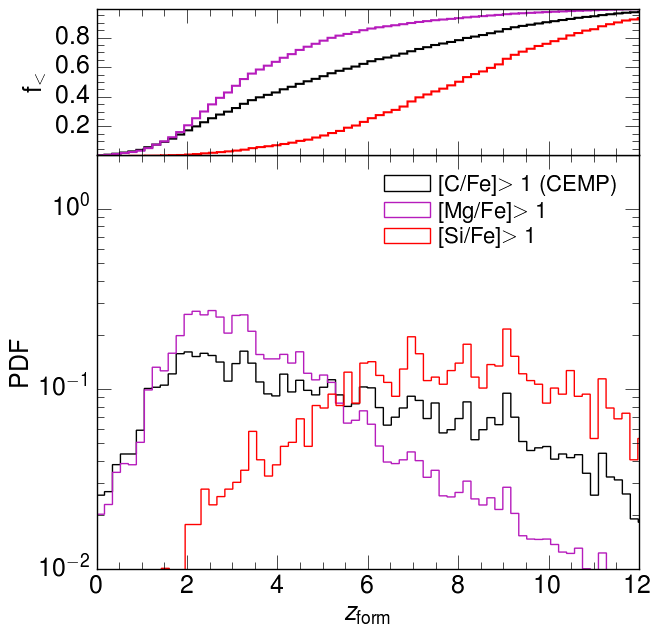}
\caption{{\em Bottom panel}: Distribution of the formation redshift of \eagle\ MW stars selected to have a give over abundance of $\alpha$-elements: [C/Fe]$>1$ (CEMP stars, {\em black line}), [Mg/Fe]$>1$ ({\em purple line}) and [Si/Fe]$>1$ ({\em red line}) The {\em top panel} shows these distributions cumulatively. }
 \label{fig_Age}
\end{figure}

To compare our predictions to data ideally involves studying the distribution of the oldest observed stars, for example in the Milky Way (MW) or its satellites. In this section we identify MW-like galaxies with central \eagle\ galaxies that inhabit a dark matter halo of mass $M_h$ in the range $10^{12}<M_h<10^{12.5}$~$h^{-1}$M$_{\odot}$ (see e.g. \citet{Wang15} and references therein for motivation of this range).

\subsection{Old but rich}
Determining the ages of stars accurately is challenging, see e.g. the review by \citet{Soderblom10} on age-dating methods, which starts with the statement that \lq ages of individual stars cannot be measured\rq. {\em Isochrone modelling} requires high-resolution and high signal-to-noise spectra, which are now available for hundreds of stars in the MW's bulge. With this method, uncertainties in the ages of individual stars are typically of the order of several gigayears or more \cite[e.g.][]{Bensby17}. {\em Nucleocosmochronometry} is based on radioactive-decay of elements and can in principle give more accurate and arguably less model dependent age estimates. This method also requires the best spectra possible, but even then the age uncertainty is of the order of gigayears; see e.g. \cite{Christlieb16} who collates data from a range of papers on ages of r-process enhanced, low-$Z$ halo stars. The abundance of O and N can be used to determine ages of red giants that agree well with those inferred from astroseismology determinations \citep{Martig16}. This promising new method still gives root mean square errors of the order of 40~per cent. Therefore even with the best data available, observations cannot really reliably determine whether an old star formed before $z=6$ (age of the Universe 1~Gyr) or $z=4$ (age = 1.6~Gyr), say.

Stellar metallicity is therefore often used in Local Group studies as a proxy for stellar age, under the reasonable assumption that on average stars that formed early on have low $Z$ \citep[e.g.][]{Frebel15}. In \eagle, an enriching particle spreads its metals to neighbouring gas particles as described by \cite{Wiersma09b}, without any further metal mixing or metal diffusion (neither physical nor numerical). We do so because at our resolution of $\sim 10^5{\rm M}_\odot$ per gas particle, we cannot hope to follow the intricacies of the turbulent metal mixing that might be occurring; see \citet{Sarmento17} for a detailed model. Nevertheless \eagle\ predicts a distribution of metallicities and metallicity gradients in MW-like galaxies that reproduces several observed trends; for example the statistics and trends of CEMP stars \citep{Sharma16b}, and the emergence of a radially flaring thin disk of relative young stars with low [$\alpha$/Fe] and a non-flaring thick disk of mostly older stars with high [$\alpha$/Fe] (see the analysis by \citealt{Navarro17} of the {\sc apostle} zooms of MW-like galaxy simulations \citep{Sawala16} performed with the \eagle\ code).

With this prescription for enrichment it is true that old stars in \eagle\ MW galaxies tend to have low [Fe/H] (Fig.~\ref{fig_FeH}): 60~per cent of stars that formed before $z=6$ have [Fe/H]$<-2$, the usual criterium for
	being classified as metal poor \cite[e.g.][]{Beers05}. However the reverse is not necessarily true: only 15~per cent of stars that are metal poor formed before $z=6$ (see also \citet{Starkenburg17}). Conversely, a small fraction of stars that formed before $z=6$ have a metallicity as high as $Z\sim 0.1Z_\odot$ - these stars are \lq old but (metal) rich\rq.

	The distribution in [Fe/H] of stars in the solar neighbourhood (defined here as stars with height $z$ above the disc of $|z|<2$~kpc, and distance $R$ from the centre in the range, $6<R/{\rm kpc}<10$, of MW type \eagle\ galaxies (blue histogram in Fig.~\ref{fig_FeH}) resembles that from observed stars from the {\sc rave} database \citep{Kunder17} (dashed blue line), peaking at solar [Fe/H] and with a rapidly decreasing fraction of stars with super-solar  [Fe/H]. However it is clear from Fig.~\ref{fig_FeH} that \eagle\ has considerably more stars with low [Fe/H] than are present in {\sc rave}, an apparent inconsistency that has been seen in previous simulations as well. \cite{Pilkington12} attributed it to a lack of metal diffusion and mixing in their simulations. Consistent with this assumption, \citet{Williamson16} show that including metal diffusion, even at low levels, suppresses the number of low metallicity stars that form in their dwarf galaxy simulation. Given this limitation in the \eagle\ simulation, we will mostly concentrate on metal {\em ratios} which are not so sensitive to lack of metal mixing, particularly when studying abundances of elements that are both produced by the same type~II SNe. We therefore explore in the next sections whether the simulation can help to  distinguish between SoRs ($z_{\rm form}>6$) and metal poor stars ([Fe/H]$<-1$) based on spatial location and abundance patterns.

\subsection{Elemental abundance patterns of SoRs}
The first generation of extremely low-$Z$ or indeed metal free stars is expected to produce elements with characteristic patterns \cite[e.g.][]{Chan16}. Low-mass stars enriched with a mixture of metals may prove to be a smoking gun for the occurrence of such SNe and enable us to study their properties. Abundances in stars \citep{Aoki14, Cooke14, Ishigaki14, Frebel15} and in low-$Z$ damped-Lyman-$\alpha$ systems (DLAs) \citep{Cooke11, Cooke17} have been interpreted as evidence for such early star formation. However, rather than examining in detail the abundance pattern associated with a given single SN, less extreme early nucleosynthesis may also leave its imprint in the abundance pattern of a large number of low-mass stars that form contemporaneously. What do we expect?
			
The standard model of stellar evolution comprises three main nucleosynthesis channels: ({\em i}) Hydrostatic burning in massive stars, $M\gtrapprox 6$~M$_\odot$, and explosive burning in their type~II core-collapse SNe descendants, which results in $\alpha$-element rich ashes (that possibly are r-process enriched \footnote{The origin of r-process elements is debated. It is possible that this process only operates in sufficiently metal rich type~II SNe, and/or it could also be that neutron star - neutron star mergers are the main source; see e.g. \cite{Thieleman17,Pian17}.}), ({\em ii}) Hydrostatic burning in intermediate mass stars ($0.5\lessapprox M/{\rm M}_\odot<6$), yielding C, O and s-process neutron capture rich ejecta when the star is on the asymptotic giant branch (AGB stars), and ({\em iii}) explosive burning in Type Ia SNe yielding Fe-rich ejecta; see e.g. \cite{Nomoto13} for a recent review.  Given the short lifetimes of massive stars, compared to that of AGB or the lower-mass progenitors of type~Ia SNe, then suggest that old stars are likely to be overabundant in $\alpha$ elements; see {\em e.g.} the discussion by \citet{Bensby14b}. If the massive stars that produce these elements are themselves metal poor, then we do not expect to detect large contributions from either the s- or the r-process. The abundance of $\alpha$ elements with higher atomic number, A, is also likely to be suppressed compared to that of lower A $\alpha$ elements, such as C, O and Mg for example (see e.g  \citealt{Sharma16b} for more discussion). CEMP-no stars (metal poor stars with [Fe/H]$\lessapprox -1$ that are carbon enhanced, [C/Fe]$\gtrapprox 1$, and without evidence for either slow or fast neutron capture elements) are therefore good candidates. This motivates us to examine $\alpha$-element abundances of \eagle\ MW-stars and of the SoRs.

Stars that formed before $z=6$ indeed are $\alpha$-enriched, with typically [Mg/Fe]$>0$ (Fig.\ref{fig_MgFe}) - consistent with enrichment by type~II ejecta. The distribution in [Mg/Fe] is broad in SoRs (and plausibly overestimated because of lack of metal mixing in \eagle\ early galaxies\footnote{A striking feature of Fig.~\ref{fig_MgFe} is the abrupt {\em lower} limit in [Mg/Fe]$\gtrapprox -1.5$, which is very apparent in SoRs (red curve) and to a much lesser extent in the PDF of all \eagle\ stars (blue curve). The origin of this lower limit is predominant enrichment by type Ia SNe. Indeed, the type~Ia yields used in \eagle\ have [Mg/Fe]$\sim -1.5$. Stars also enriched with Fe by type~II SNe have higher values of [Si/Fe].}). Comparing the blue and red curves suggests that selecting stars with [Mg/Fe]$>1$, say, could be used to identify SoRs. To examine this in more detail we plot the PDF of formation redshifts for stars with [$\alpha$/Fe]$>1$ in Fig.\ref{fig_Age}, for $\alpha$ elements C, Mg, and Si. From this we see that 80\% of stars with [C/Fe]$>1$ or [Mg/Fe]$>1$ formed before $z\sim 2$, whereas 80\% of stars with [Si/Fe]$>1$ even formed before $z=6$. This would suggest that a high value of [Si/Fe] is therefore an even better indicator that the star is very old. 
	
We characterise a selection criterion $S$ ({\em e.g.} [C/Fe]$>1$) for SoRs in terms of its \lq purity\lq, $P$, and \lq completeness\rq, $C$, defined as 
\begin{eqnarray}
P &=& {{\cal P}(S \bigcap {\rm SoR})\over {\cal P}(S)}\nonumber\\
C &=& {{\cal P}(S \bigcap {\rm SoR}) \over {\cal P}({\rm SoR})}\,.
\label{eq:PC}
\end{eqnarray}
Here, ${\cal P}({\rm SoR})$ denotes the fraction of stars that are SoRs, ${\cal P}(S\bigcap {\rm SoRs})$ the fraction of stars that satisfy $S$ and are SoRs, {\em etc}. A pure selection criterion has $P\sim 1$, meaning a star that satisfies $S$ is most likely a SoR, a complete selection criterion has $C\sim 1$, meaning almost all SoRs satisfy it. For given $S$, we can compute $P(S)$ and $C(S)$, and varying the value of $S$ yields a curve in a diagram of $P$ versus $C$. In Fig.~\ref{fig_PC}, we plot these curves for [C/Fe], [Mg/Fe] and [Si/Fe] obtained for all the SoRs in MW-like galaxies in \eagle\ (thin solid lines). At face value, these curves suggest that [Si/Fe] is the best compromise between purity and completeness, since, for example, [Si/Fe]=$0.7$ yields $P\sim 0.7$ ({\em i.e.} only 30~per cent of stars with [Si/Fe]$>0.7$ are {\em not} SoRs) and $C\sim 0.3$ ({\em i.e.} 30~per cent of SoRs also have [Si/Fe]$>0.7$). 

We attempt to compare these trends in \eagle\ to those measured in observed very low metallicity stars taken from the {\sc saga} database \citep{Suda11} \footnote{The observational data represent a diverse collection of abundances of metal poor stars collected from the literature, and this comparison data is therefore neither complete nor unbiased - see \citet{Suda11} for details.}. To do so we assume that all {\sc saga} stars {\em are} SoRs, which is not unreasonable given their very low [Fe/H]. This allows us to compute the completeness, $C$, for {\sc saga} stars. Of course we cannot compute their $P$ values, since this requires knowledge of the value of $S$ for all MW stars, and in addition which of these is a SoR. We therefore use the value of $P$ measured from \eagle\ instead; this results in the thick solid lines in Fig.~\ref{fig_PC}.
	It is immediately clear that for a given abundance ratio, [Mg/Fe] and [Si/Fe] are much more complete in \eagle\ than in {\sc saga}. We discuss this apparent discrepancy in more detail in Appendix~\ref{sect:Appendix1}. However \eagle\ does reproduce the dependence of $C$ on [C/Fe] measured in {\sc saga} well: the fraction of SoRs stars that have [C/Fe] greater than some value agrees relatively well. For example the fraction of SoRs with [C/Fe] greater than 0.6, 0.8 and 1.0 in \eagle\ is 17, 12 and 10~per cent, respectively, whereas when using {\sc saga} (for computing completeness) it is slightly higher at 26, 20 and 17~per cent.

In \eagle, high values of [C/Fe] select SoRs with high purity and completeness: a large fraction of CEMP stars are SoRs, and vice-versa. Given the relative good agreement in [C/Fe] ratios measured for SoRs in the {\sc saga} database, suggests that the same is true for MW stars.

%}

\begin{figure}
 \centering
 \includegraphics[width=\linewidth]{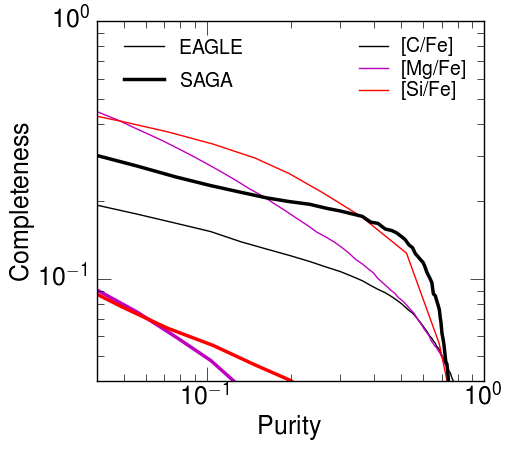}
\caption{Purity $P$ versus completeness $C$, as defined in Eq.~(\ref{eq:PC}) for [C/Fe] ({\em black lines}), [Mg/Fe] ({\em purple lines}) and [Si/Fe] ({\em red lines}); {\em thin solid lines} are for \eagle\ MW stars, {\em thick solid lines} when the completeness is calculated using stars from the {\sc saga} data base \citep{Suda11}. Circles from left to right correspond to overabundances of [$\alpha$/Fe]$=0.6,0.8,1.0$ for each element. 
}
 \label{fig_PC}
\end{figure}
The approximate correspondence between CEMP stars and SoRs is illustrated in more detail in Fig.~\ref{fig_CFeH}, where we plot the fraction of stars that are carbon enhanced as a function of [Fe/H]. This fraction {\em increases} with {\em decreasing} metallicity,  rising from 10~per cent for stars with [Fe/H]$<-2$ to $\approx60$~per cent for stars with [Fe/H]$<-4$ (blue solid curve). The trend in \eagle\ agrees strikingly well with the observational data from \cite{Lee13} (blue points with error bars)\footnote{The upcoming revised numbers from the observations are expected to be higher (Tim Beers pvt. comm. and \citealt{Yoon18}).}, which may be fortuitous. If we restrict the analysis to stars that formed before $z=6$ (SoRs),  then the fraction is $\approx20$~percent at [Fe/H]$=-2$ and $\approx60$~percent at [Fe/H]$=-4$. This demonstrates once more that a large fraction of SoRs are CEMPs, particularly at the lowest metallicities. Vice versa, the purple lines show that a large fraction of CEMPs are SoRs:  40~per cent of stars with [Fe/H]$<-2$ and [C/Fe]$>1$ formed before $z=6$ in \eagle\, rising to 60~per cent for [Fe/H]$<-4$. We conclude that, at low metallicities ([Fe/H]$<-4$), to a good approximation, {\em CEMP stars and SoRs are the same population}.

\begin{figure}
 \centering
 \includegraphics[width=\linewidth]{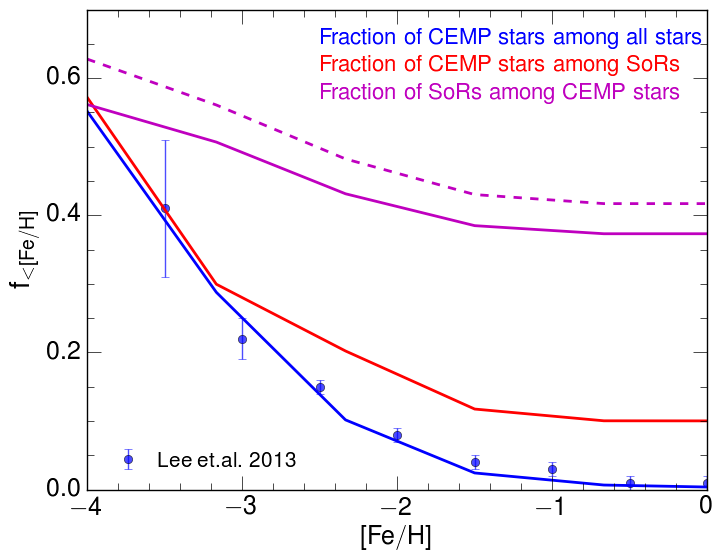}
\caption{
	Fraction of stars with [C/Fe]$>1.0$ and [Fe/H]$<-1$ ($f_<$) as a function of their [Fe/H], for \eagle\ MW stars ({\em blue solid line}), and for \eagle\ SoRs ({\em red solid line}). {\em Blue data points with error bars} show $f_<$ for observed MW stars taken from \citet{Lee13}. The good agreement with the corresponding fraction in \eagle\ is striking. The fraction of \eagle\ CEMP stars  ([C/Fe]$>1.0$, [Fe/H]$<-1$), that formed before
	$z=6$ ($z=5.5$) as a function of [Fe/H], is plotted as a ({\em solid (dashed) purple line}).}
 \label{fig_CFeH}
\end{figure}

%%%%%%%%%%%%%%%%%%%%%%%%%%%%%%%%%%%%%%%%%%%%%%%%%%%%%%%%%%%%%%%%%%%%%%%%%%%%%%%
\subsection{Spatial distribution of SoRs and CEMP stars in the Milky Way}
%%%%%%%%%%%%%%%%%%%%%%%%%%%%%%%%%%%%%%%%%%%%%%%%%%%%%%%%%%%%%%%%%%%%%%%%%%%%%%%
\label{sect:MWSoRs}
\begin{figure}
 \centering
 \includegraphics[width=0.9\linewidth]{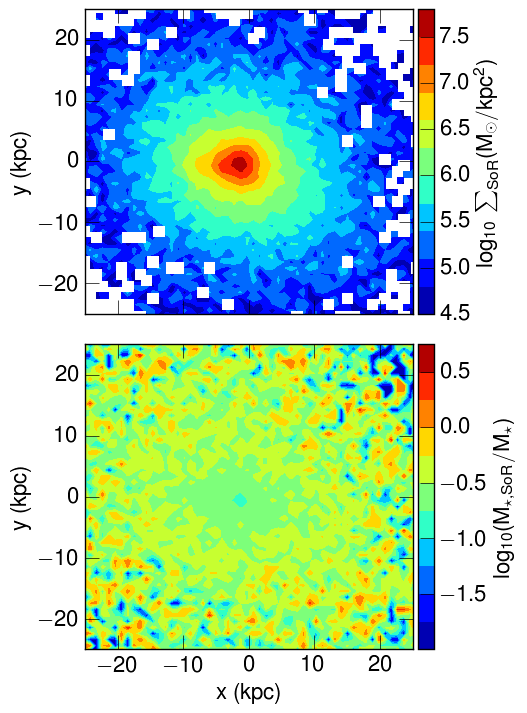}
 \includegraphics[width=0.9\linewidth]{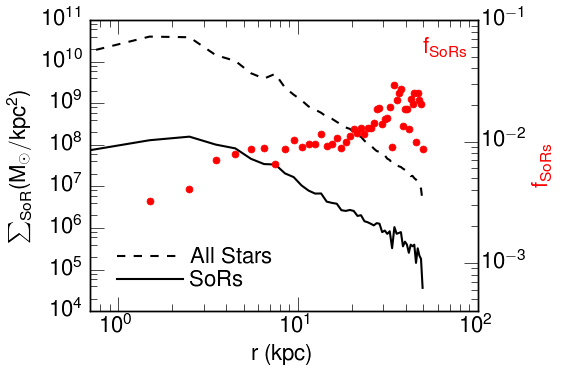}
 \caption{Spatial distribution of SoRs in \eagle\ MW galaxies. {\em Top panel:} surface density of SoRs (stars that formed before $z=6$) in MW discs seen face-on. Colours encode stellar surface density. {\em Middle panel:} ratio of surface density of SoRs over total stellar surface density. These plots show that SoRs are centrally concentrated, but less so than stars that form later, such that the fraction of SoRs {\em increases} with distance $r$ from the centre. {\em Bottom panel:} radial profile of stellar surface density for SoRs ({\em solid black line}), and for all stars ({\em black dashed line}). {\em Red dots} show the fraction of SoRs (right hand $y$-axis).}
 \label{fig_radface}
\end{figure}
%%%%%%%%%%%%%%%%%%%%%%%%%%%%%%%%%%%%%%%%%%%%%%%%%%%%%%%%%%%%%%%%%%%%%%%%%%%%
Observational searches for CEMP stars often scour the stellar halo (by studying halo stars that pass close to the Sun), and indeed meet with considerable success as discussed in the Introduction. Here we examine the spatial distribution of both SoRs and CEMPs in \eagle\ MW galaxies.

We find that the distribution of SoRs is centrally concentrated, reaching a central surface density of $\sim 10^8$~M$_\odot$~kpc$^{-2}$. However stars that form more recently are even more centrally concentrated, so that the fraction of stars that are SoRs {\em increases} with distance from the centre (Fig.~\ref{fig_radface}). At the location of the Sun, $r\sim 8~$kpc, we predict that $\approx 1$~per cent of stars within $\lvert z \rvert<2$~kpc of the disc are SoRs. So even though the number density of SoRs is higher in the bulge, it might still be easier to find them in the Solar neighbourhood.

These findings agree with earlier work. \cite{White00} used dark matter only simulations, labelling some of them as a \lq first stars\rq, and found that such old stars are concentrated in the bulge. \cite{Brook07} confirm these findings but also report the possibility of there being a population of metal poor stars in the Galactic halo. \cite{Tumlinson10}  and the semi analytic work by \cite{Salvadori10}, also conclude that the central region of the galaxy hosts most of its metal poor stellar population. The simulation of the MW-Andromeda \lq Local Group\rq, analysed by \cite{Starkenburg17}, that has the same subgrid physics as in this study, also predicts that most of the metal poor stars are centrally concentrated. In Fig.~\ref{fig_radface} we selected stars by age rather than metallicity, however as we discussed previously, more than 50~per cent of stars that are metal poor, [Fe/H]$<-2$, are also old, formation redshift $z>6$. So the good agreement between our findings and those of other groups is not surprising.

We now turn our attention to the spatial distribution of CEMP stars. CEMP stars are further divided into CEMP-s stars (which are overabundant in slow neutron capture (\lq s-process\rq) elements such as barium), CEMP-r stars (which are overabundant in 
rapid neutron capture (\lq r-process\rq) elements), and CEMP-no that are not overabundant in either neutron capture element, and CEMP-rs that are overabundant in both; \citet[see e.g.][]{Beers05}. The origin of CEMP-s stars is typically attributed to mass transfer from an AGB binary companion, but \cite{Sharma17}  suggests they may also originate from star formation in gas enriched only (or at least mostly) by ejecta from AGB stars. In contrast, CEMP-no stars may result from enrichment by low-$Z$ type~II SNe. Observationally, CEMP-no stars have typically lower [Fe/H] than CEMP-s stars \citep{Aoki07}: at [Fe/H]$<-3$ most observed CEMP stars are of type CEMP-no \citep{Frebel15}. Interestingly, the fraction of CEMP-no stars {\em increases} with distance $r$ from the galaxy's centre \citep{Carollo14}. What can be the reason for this?
%%%%%%%%%%%%%%%%%%%%%%%%%%%%%%%%%%%%%%%%%%%%%%%%%%%%%%	
\begin{figure}
 \centering
\includegraphics[width=\linewidth]{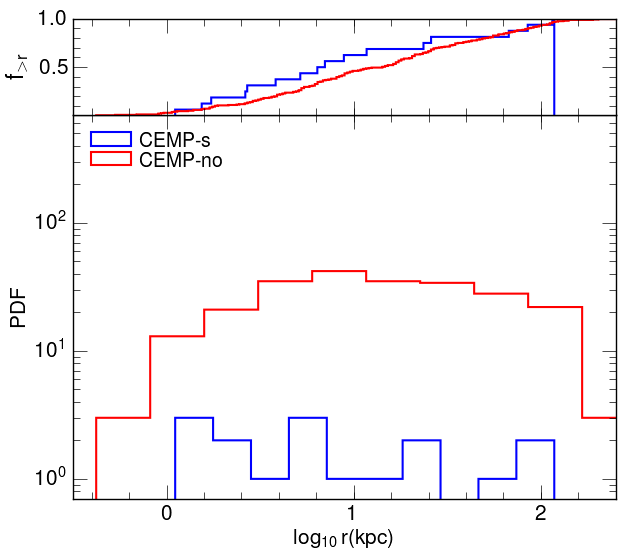}
\caption{{\em Bottom panel:} radial distribution of CEMP stars ([C/Fe]$>1$, [Fe/H]$<-1$) in a galaxy like the Milky Way selected from \eagle\ for type CEMP-s ({\em blue line}, A(C)$>7.1$), and type CEMP-no ({\em red line}, A(C)$<7.1$)). {\em Top panel:} corresponding cumulative distribution.}
 \label{fig_cemps}
\end{figure}
%%%%%%%%%%%%%%%%%%%%%%%%%%%%%%%%%%%%%%%%%%%%%%%%%%%%%%%

In \eagle\ we found that a considerable fraction of SoRs are CEMPs, and hence we expect the fraction of CEMP stars to increases with distance $r$ from the centre, and with height $z$ above the disc, as do SoRs. We further divide CEMP stars ([C/Fe]$>1$, [Fe/H]$<-1$) in \eagle\ into subtypes by using their absolute carbon abundance, A(C), and select CEMP-s stars by requiring  A(C)$>7.1$, and CEMP-no stars by requiring  A(C)$<7.1$, as in \citet{Yoon16}. In Figure~\ref{fig_cemps} we plot the radial distribution of CEMP-no and CEMP-s stars in \eagle\ MW galaxies. We find that CEMP-no stars have a more extended distribution compared to the CEMP-s stars: 50~per cent of CEMP-no stars have $r\lessapprox13~$kpc, whereas 50~per cent of CEMP-s stars have $r\lessapprox7$~kpc; at $r>10$~kpc most of the stars CEMP stars are CEMP-no. This is remarkably similar to observed, \citep{Carollo14, Lee17}. For example, \cite{Carollo14} find that 70~percent of stars in the outer halo  are of CEMP-no subtype.

The reason that most CEMP stars are CEMP-no in the outskirts of the {\sc eagle} MW galaxies is closely related to the nature of star formation in progenitors of MW galaxies, as discussed by \cite{Sharma16, Sharma17}. Star formation in these dwarf galaxies is very bursty leading to  poor mixing of type~II SN and AGB ejecta - this is what gives rise to the CEMP-no and CEMP-s stars in the first place, with CEMP-no stars forming during blow-out of a galaxy. When such galaxy fragments merge during the hierarchical build-up of the MW, many of their collisionless stars do not end-up in the centre of the merger remnant but  rather in its outskirts: most of the extended population of CEMP-no stars was accreted (formed {\em ex situ}), while the majority of the centrally concentrated CEMP-s and CEMP-no stars were formed later and {\em in situ} (see Fig.~\ref{fig_cumR} and \ref{fig_why}). 
%%%%%%%%%%%%%%%%%%%%%%%%%%%%%%%%%%%%%%%%%%%%%%%%%%%%%
\subsection{SoRs in the integrals of motion plane}
%%%%%%%%%%%%%%%%%%%%%%%%%%%%%%%%%%%%%%%%%%%%%%%%%%%%%
\begin{figure}
 \centering
 \includegraphics[width=\linewidth]{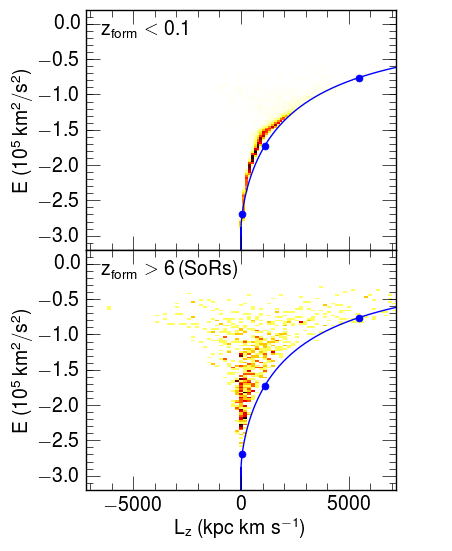}
 \caption{Distribution of \eagle\ MW stars in the integral of motion plane. Pixels are coloured by the density of stars in $E-L_z$ space, for recently formed stars ({\em top panel}) and SoRs stars ({\em bottom panel}), selected from an \eagle\ MW galaxy. The {\em blue curve} is the relation between $E$ and $L_z$ from Eq.~\ref{eq:EM}, for particles on a circular orbit, in an NFW halo with scale radius $r_s\approx5$~kpc and circular velocity at virial radius $v_{200}\sim200$~km~s$^{-1}$; the blue circles, from the lowermost to uppermost denote values corresponding to $x=0.1,1$ and $5$, respectively.}
 \label{fig_ezlz}
\end{figure}
%%%%%%%%%%%%%%%%%%%%%%%%%%%%%%%%%%%%%%%%%%%%%%%%%%%%%
The spatial distribution of SoRs or CEMPs is usually described in terms of their distance to the centre and their height above the disc, $(r,z)$, but of course these are not constants of the motion. A better description is in terms of their energy per unit mass, and spin angular momentum about the symmetry axis of the galaxy, $(E,L_z$), which are at least approximately conserved.
	This distribution is plotted in Fig.~\ref{fig_ezlz} as a 2D histogram of the density of stars with given $E$ and $L_z$, for stars in one \eagle\ MW-galaxy.

Not surprisingly, the distributions of old and young stars is strikingly different: recently formed stars (top panel) hug the blue line, which corresponds to stars that are on nearly circular orbits in the disc, whereas old stars (bottom panel) lie almost symmetrically in the funnel-shaped region delineated by the (same) blue line on the right, and its mirror with respect to $L_z=0$ on the left. Clearly disc stars and accreted stars can be distinguished easily in the $E-L_z$ \lq integrals of motion\rq\ plane \citep[e.g.][]{Helmi17}.

It is useful to make a simple model that captures where disc stars ({\em i.e.} stars on a circular orbit) and halo stars are located in an $E-L_z$ plane \citep[e.g.][]{Sellwood02}. We begin by writing $L_z=v_c\,r$, in terms of the velocity $v_c$ of a star on a circular orbit with radius $r$, $v_c^2={\rm G}M/r$, where $M$ is enclosed mass, and $E=(1/2)v_c^2+\phi$; $\phi$ is the gravitational potential. Once a potential-density pair is given, $E(L_z)$ can be computed.

In the particular case of a Navarro, Frenk and White (NFW) halo \citep{Navarro96}, with scale radius $r_s$ and maximum circular velocity $v_{c, {\rm max}}$, this relation can be written in parametric form as
%\begin{eqnarray}
%	{L_z\over a\,E_0^{1/2}} &=& {c^{3/2}\over 1+c}\nonumber\\
%    {E\over E_0}      &=& -1 + {1+c/2\over (1+c)^2)}\,,
%\end{eqnarray}
\begin{eqnarray}
	{L_z\over r_s\,v_s} &=& x\left({\ln(1+x)\over x}-{1\over 1+x}\right)^{1/2} \nonumber\\
    {E\over v_s^2}      &=& -{1\over 2}\left({\ln(1+x)\over x}+{1\over 1+x}\right)\nonumber\\
	{L\over r_s\,v_s}   &\approx & {4\sqrt{6}\over 9}\,(1+{E\over v_s^2})^{3/2};\, \, \, \, r\ll r_s\,.
	\label{eq:EM}
\end{eqnarray}

where $x\equiv r/r_{\rm s}$. For a given energy $E$, the angular momentum of any star is $|L|\leq L_z$, with $L_z$ given by Eq.~(\ref{eq:EM}), since a circular orbit has the largest angular momentum. Any star in this spherical halo therefore lies in the funnel-shaped region, defined by the curves $L_z(E)$ and $-L_z(E)$. The third equation is the first non-zero term in  a Taylor expansion close to the centre, $x=r/r_{\rm s}\ll 1$. In these relations, $v_s$ depends on $v_{200}=({\rm G} M_{200}/r_{200})^{1/2}$ and the concentration, $c$, of the NFW halo, as
\begin{equation}
v_{\rm s} = v_{\rm 200}\,({\ln(1+c)\over c} - {1\over 1+c})^{-1/2}\,.
\end{equation}
We used common notation, in which the radius of the halo is $r_{200}$, defined such that the mean density within $r_{200}$ is 200 times the critical density, and $M_{200}$ is the total mass enclosed within $r_{200}$.

The parametric relation between $E$ and $L_z$ of Eq.~({\ref{eq:EM}) is plotted in Fig.~\ref{fig_ezlz} (blue curve), taking
	$c=10$ (which yields $v_s\approx 2.6\,v_{\rm 200}$), $v_{200}\approx200$~km~s$^{-1}$  and $r_{\rm s} \approx 5$~kpc (reasonable values for a MW-like halo). It captures well the edges of the allowed funnel shaped region occupied by stars.

\section{S\lowercase{o}R\lowercase{s} in other galaxies}
\subsection{Distribution of SoRs across the galaxy population}
\label{abund_gal}
\label{sec_where}
The shape of the galaxy stellar mass function at redshift $z=0$ is well described by a Schechter function \citep{Schechter76}, {\em i.e.} a power-law at low mass and an exponential cut-off above a characteristic mass. Since the galaxy stellar-mass function is part of the calibration of \eagle's sub-grid parameterisation, the simulation reproduces the observed galaxy stellar mass function well \citep{Schaye15}. For such a function, most mass is in massive galaxies, in particular we find that in \eagle\
	more than 70~per cent of the total stellar mass is in galaxies with $M_\star\gtrapprox 10^{10}$~M$_\odot$ (blue line in Fig.~\ref{fig_abund}). 

The distribution of SoRs among those galaxies (black line in the lower panel of Fig.~\ref{fig_abund}) is not very different	
	from that of total mass for massive galaxies, but there is a noticeable difference at the low mass end: 20~per cent of SoRs
	reside in galaxies with $M_\star\lessapprox 10^{9.2}$~M$_\odot$, whereas for all stars that limit is $M_\star\lessapprox 10^{9.8}~$M$_\odot$: a larger fraction of SoRs is contained in lower-mass galaxies compared to the total stellar mass.
	Even though most SoRs are hosted by massive galaxies (70~per of the total mass in SoRs is in galaxies with $M_\star\gtrapprox 10^{10}$~M$_\odot$), the fraction of mass in SoRs in these galaxies is typically quite low, of order 0.03~per cent.
	Vice versa, a small fraction of the total SoRs is in low-mass galaxies, yet some of these can have as much as 10~per cent of their total mass in SoRs. This is not contradictory, because a large fraction of low-mass galaxies do not host any SoRS at all: at $M_\star=10^8$~M$_\odot$, as much as 6~per cent of galaxies do not host any SoRs. We find that there is no obvious dependence
	of the SoRs mass fraction on whether a galaxy is a central or a satellite. 
%%%%%%%%%%%%%%%%%%%%%%%%%%%%%%%%%%%%%%%%%%%%
\begin{figure}
 \centering
 \includegraphics[width=\linewidth]{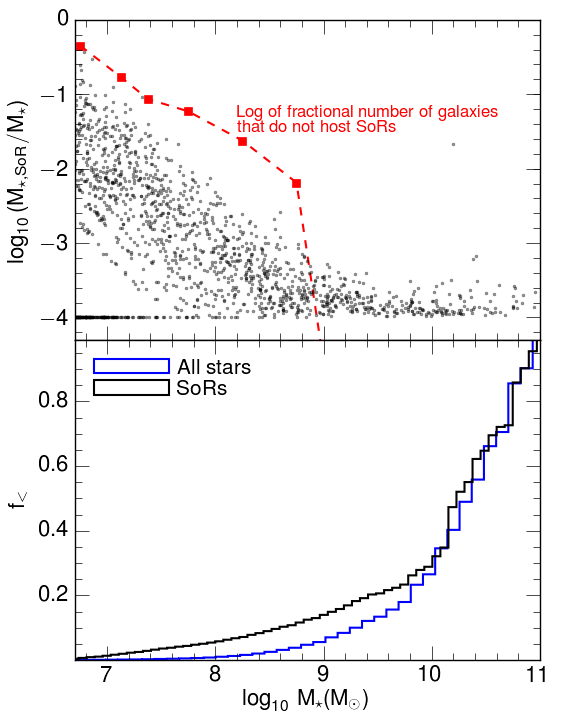}
 \caption{The distribution of siblings of reionizers (SoRs) across galaxies in \eagle\ simulation L0025N0752 at redshift $z=0$.
	{\em Top panel}: stellar mass fraction in SoRs, $f_{\rm SoRs}=M_{\star, {\rm SoR}}/M_\star$, as a function of $M_\star$: every black dot corresponds to an \eagle\ galaxy, those without SoRs are arbitrarily placed at $f_{\rm SoRs}=10^{-4}$.
	The {\em red dashed line} is the logarithm of the fraction of galaxies that do not host any SoRs. {\em Bottom panel:} cumulative mass fraction in SoRs in galaxies below a given $M_\star$ ({\em black line}), the cumulative mass fraction for all stars is plotted as a {\em blue line}.}
\label{fig_abund}
\end{figure}
%%%%%%%%%%%%%%%%%%%%%%%%%%%%%%%%%%%%%%%%%%%%%%

How do these results compare to previous findings? From their simulations, \citet{Gnedin06} conclude that low mass dwarf satellites of the Milky Way may contain a high fraction of SoRs \citep[see also][]{Madau08}, with 5-15 per
	cent of the objects that exist at $z=8$ surviving to the present without undergoing significant evolution. Consistent with this, \citet{Frebel14} reported that the nearby dwarf satellite Segue-I  contains an unusually high fraction of metal poor stars (with all seven observed stars metal poor, [Fe/H]$<$-1.4, and $\alpha$-enhanced, [$\alpha$/Fe]$>$0.5); see also \citet{Webster16}. This is not inconsistent with our findings, however we suggest that the {\em scatter} in the abundance of SoRs at low values of $M_\star$ may be very large.

\subsection{Spatial distribution of SoRs in galaxies}
The radial profile of total stellar mass is compared to that of SoRs in Figure \ref{fig_cumR}, for three bins in halo mass.
	There is a clear trend: SoRs are less centrally concentrated than the total stellar distribution, and this is more so
	in more massive halos. We demonstrated that the distribution of SoRs is more extended than that of all stars in MW-like galaxies 
	in section~\ref{sect:MWSoRs}, clearly this is true for all galaxies. There we argued that this is because many SoRs are
	accreted, building up an extended halo of very old stars. 

The fraction of SoRs in a galaxy that are accreted, rather than formed in situ, increases with halo mass, $M_h$ (Fig.~\ref{fig_why}, {\em bottom panel}). For a MW-like halo mass of $M_h\sim 10^{12}$~M$_\odot$ we find that 80~per cent of SoRs are accreted, and this fraction rises further with increasing $M_h$. However, there is considerable scatter. At low values of $M_h$, the accreted fraction drops quickly, to as low as 20~per cent at $M_h\sim 10^{10.2}$~M$_\odot$.
	
The relation between the mass of the halo in which a SoR formed, and the mass of the halo in which it resides today, is illustrated further in Fig.~\ref{fig_why}, {\em top panel}), which shows the fraction of mass in SoRs that resides in halos of a given mass.
	At high-$z$, SoRs reside predominantly in relatively low halos, $M_h\lessapprox 10^{10.5}$ and $\lessapprox 10^{11}$~M~$_\odot$ at $z=8$, and $z=6$, respectively. Therefore SoRs form in relatively low mass galaxies. Yet today, they are found predominantly in massive halos, as we showed before. This is consistent with these massive halos having mostly accreted their SoRs, which explains why their spatial distribution is so much more extended than that of the in situ formed stars, which dominate the stellar mass of a galaxy.

From the above we conclude that SoRs formed in low-mass halos. When these halos merged hierarchically to form more massive halos, these stars were accreted as well and build-up the stellar halo of the merger remnant. As a result, SoRs at $z=0$, inhabit mostly the outskirts of the more massive halos.

In fact, in more massive haloes, $M_{\rm h}>10^{13}{\rm M}_\odot$, which includes groups and clusters, approximately 80~percent of the SoRs reside in the outskirts of the central galaxy at distances $r>20$~kpc ({\em magenta} curve in Fig.~\ref{fig_cumR}). Furthermore, 3~percent of all the stars at $r>20$~kpc in such massive haloes are SoRs. Combined with the fact that about 50~per cent of SoRs are associated with these massive galaxies (Fig.~\ref{fig_abund}), we conclude that almost half of all SoRs in the Universe form part of the stellar halo of massive galaxies, where they contribute to their intra-cluster light.

\begin{figure}
 \centering
 \includegraphics[width=\linewidth]{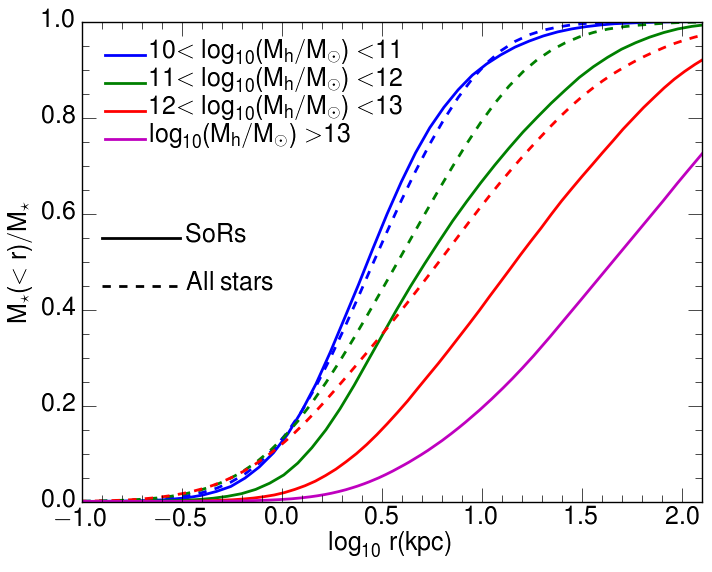}
 \caption{Cumulative stellar mass distribution as a function of distance, $r$, to the centre of the galaxy for \eagle\ galaxies at redshift $z=0$. Colours refer to galaxies in a narrow bin of halo mass: {\em blue lines} $10^{10}\leq M_h/{\rm M}_\odot\leq 10^{11}$, {\em green lines} $10^{11}\leq M_h/{\rm M}_\odot\leq 10^{12}$, and {\em red lines} $10^{12}\leq M_h/{\rm M}_\odot\leq 10^{13}$. The distribution for all stars is plotted as dashed lines, that of SoRs as solid lines. The distribution of SoRs in halos with $M_h/{\rm M}_\odot\geq 10^{13}$ selected from the \eagle\ simulation Ref-L100N1504,
	which has co-moving volume (100~Mpc)$^3$, is shown as the {\em magenta} curve. At any halo mass, SoRs are relatively more abundant in galaxy outskirts; this is particularly evident for the most massive halos ({\em red} and {\em magenta} curves).}
 \label{fig_cumR}
\end{figure}
\begin{figure}
 \centering
 \includegraphics[width=\linewidth]{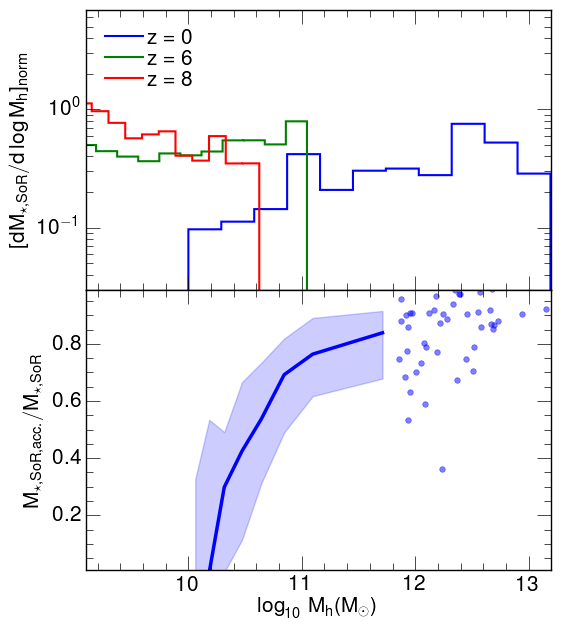}
\caption{{\it Upper panel}: Distribution of SoRs in haloes in \eagle\ at redshift 0 (blue), 6 (green) and 8 (red). Most of the SoRs at high redshift live in low mass haloes contrary to the trend at low redshift. {\it Lower Panel}: the median fraction of SoRs  that were accreted via mergers post reionization at $z\le6$ in present day galaxies is shown as a solid blue curve with shaded region showing 20th and 80th percentile. There are few galaxies in the simulation volume with $M_h>7\times10^{11}$~M$_\odot$: these are shown individually as {\em blue circles}.}
 \label{fig_why}
\end{figure}

\section{Summary and Conclusion}
\label{sec_sum}
The Universe was most likely reionized by massive stars. These have short lifetimes, and hence leave no age-datable remnants today. However any low-mass stars that formed contemporaneously with these \lq reionizers\rq\ can be studied in the local universe and the galactic archaeology of such \lq siblings of reionization\rq\ (SoRs) is an alternative way to study reionization.

In this paper, we identified SoRs in the \eagle\ simulations \citep{Schaye15, Crain15} as stars that formed before redshift $z=6$. This is motivated by the model for reionization by \cite{Sharma17}, who suggest that stars formed at high surface density of star formation are mostly responsible for producing the ionising photons that escaped their galaxy and ionised the intergalactic medium. Their reasoning is that, when star formation occurs at high surface density, it efficiently drives winds that create channels through which ionising photons can escape. In the \eagle\ simulations we find that most star formation above $z=6$ occurs at the required threshold surface density - because high-$z$ galaxies form stars vigorously and are small - hence our identification of SoRs with stars that form before $z=6$ (none of our results would change significantly if we used $z=5$ or $z=7$ instead).

We examined the properties of SoRs in \eagle\ galaxies that resemble the Milky Way (MW) in Section~3, assuming MW galaxies are hosted in halos of mass $M_h\sim 10^{12}$~M~$_\odot$. We find that SoRs are significantly overabundant in $\alpha$-elements, a consequence of rapid enrichment by low-$Z$ type~II SNe. Comparing these overabundances to MW stars from the {\sc saga} database \citep{Suda11}, we find that overabundances in Si and Mg are higher in \eagle\ than observed, plausibly as a result of the lack of mixing with type~Ia ejecta and overly high type~II yields in the simulation. This does not affect carbon as much because carbon is not produced in type~Ia SNe. Indeed, we find good agreement between [C/Fe] as a function of [Fe/H] between \eagle\ and {\sc sage}.

Examining [C/Fe] abundances in \eagle\ MW-galaxies in more detail, we find that a significant fraction of SoRs are CEMP stars and similarly, a large fraction of CEMP stars, in particular of CEMP-no stars (lacking in neutron capture elements), are SoRs. Quantitatively, 45~per cent of CEMP stars ([C/Fe]$>1$) at [Fe/H]$<-2$  are SoRs, and conversely 20~per cent of SoRs at [Fe/H]$<-2$ are CEMP stars (with [C/Fe]$\gtrapprox 1$). These numbers may well depend on the physical models implemented in the simulation; nevertheless the trends that we obtain are likely robust.

One of the trends that we find is that the fraction of stars (and SoRs) that are carbon enhanced (CEMP) increases rapidly with decreasing metallicity (Fig.~\ref{fig_CFeH}). At metallicities of [Fe/H]$<-4$, more than half of CEMP stars are SoRs, and conversely, a similar fraction ($\gtrsim50$~per cent) of SoRs are CEMP stars. Furthermore, the CEMP stars at such low metallicities are of CEMP-no subtype. This leads us to conclude that, at lowest metallicities, to a good approximationCEMP-no stars and SoRs {\em are the same population}. 

We also examined the distribution of SoRs in other galaxies in the simulation (Section~4). Some low-mass galaxies have no SoRs at all, whereas some have a very high fraction of SoRs, up to 10~per cent. More massive galaxies typically have a small fraction of SoRs, of order 0.01~per cent. The shape of the galaxy stellar mass function still results in most SoRs inhabiting massive galaxies: approximately 70~per cent of SoRs are in galaxies of mass $M_\star\gtrapprox 10^{10}$~M$_\odot$. A large fraction of these are accreted rather than formed in situ. The net result is that SoRs tend to inhabit the {\em outskirts} of massive galaxies.

We therefore conclude that CEMP-no stars are the siblings of the stars that reionized the Universe. These stars formed in small galaxies before $z=6$ and $\gtrsim 50$~percent of them ended up in the outskirts of massive galaxies today, where they contribute to the intracluster light. Determination of the mass in such intracluster stars can be translated into constraints on the initial mass function of the stars that reionized the Universe since these low mass stars formed contemporaneously with the high mass stars that produced the ionizing photons.  Obtaining more details on the abundances of CEMP-no stars in the intracluster light may be challenging, but a good start could be made by studying them in the Milky Way, where we predicted that SoRs (and CEMP-no stars) are mostly found in the stellar halo. Our model of reionization would be invalidated if such investigations were to uncover a very low overall number of such stars.

\section*{Acknowledgements}
We thank an anonymous referee for thoughtful comments. We thank our colleagues (J. Schaye, M. Schaller, R. Crain and R. Bower) for sharing with us the data from the \eagle\ simulation. This study was funded by the Science and Technology Facilities Council [grant number ST/F001166/1] and by the Interuniversity Attraction Poles Programme of the Belgian Science Policy Office ([AP P7/08 CHARM]). The DiRAC Data Centric system at Durham University was used this study. This system is run by the Institute for Computational Cosmology on behalf of the STFC DiRAC HPC Facility (www.dirac.ac.uk); the equipment was funded by BIS National E-Infrastructure capital grant ST/K00042X/1, STFC capital grant ST/H008519/1, and STFC DiRAC; as a part of the National E-Infrastructure.  M.~S. is supported by an STFC post-doctoral fellowship.

\footnotesize{\bibliography{ref_where}}

\begin{thebibliography}{}
\makeatletter
\relax
\def\mn@urlcharsother{\let\do\@makeother \do\$\do\&\do\#\do\^\do\_\do\%\do\~}
\def\mn@doi{\begingroup\mn@urlcharsother \@ifnextchar [ {\mn@doi@}
  {\mn@doi@[]}}
\def\mn@doi@[#1]#2{\def\@tempa{#1}\ifx\@tempa\@empty \href
  {http://dx.doi.org/#2} {doi:#2}\else \href {http://dx.doi.org/#2} {#1}\fi
  \endgroup}
\def\mn@eprint#1#2{\mn@eprint@#1:#2::\@nil}
\def\mn@eprint@arXiv#1{\href {http://arxiv.org/abs/#1} {{\tt arXiv:#1}}}
\def\mn@eprint@dblp#1{\href {http://dblp.uni-trier.de/rec/bibtex/#1.xml}
  {dblp:#1}}
\def\mn@eprint@#1:#2:#3:#4\@nil{\def\@tempa {#1}\def\@tempb {#2}\def\@tempc
  {#3}\ifx \@tempc \@empty \let \@tempc \@tempb \let \@tempb \@tempa \fi \ifx
  \@tempb \@empty \def\@tempb {arXiv}\fi \@ifundefined
  {mn@eprint@\@tempb}{\@tempb:\@tempc}{\expandafter \expandafter \csname
  mn@eprint@\@tempb\endcsname \expandafter{\@tempc}}}

\bibitem[\protect\citeauthoryear{{Amor{\'{\i}}n} et~al.,}{{Amor{\'{\i}}n}
  et~al.}{2017}]{Amorin17}
{Amor{\'{\i}}n} R.,  et~al., 2017, \mn@doi [Nature Astronomy]
  {10.1038/s41550-017-0052}, \href
  {http://adsabs.harvard.edu/abs/2017NatAs...1E..52A} {1, 0052}

\bibitem[\protect\citeauthoryear{{Aoki}, {Beers}, {Christlieb}, {Norris},
  {Ryan}  \& {Tsangarides}}{{Aoki} et~al.}{2007}]{Aoki07}
{Aoki} W.,  {Beers} T.~C.,  {Christlieb} N.,  {Norris} J.~E.,  {Ryan} S.~G.,
  {Tsangarides} S.,  2007, \mn@doi [\apj] {10.1086/509817}, \href
  {http://adsabs.harvard.edu/abs/2007ApJ...655..492A} {655, 492}

\bibitem[\protect\citeauthoryear{{Aoki}, {Tominaga}, {Beers}, {Honda}  \&
  {Lee}}{{Aoki} et~al.}{2014}]{Aoki14}
{Aoki} W.,  {Tominaga} N.,  {Beers} T.~C.,  {Honda} S.,   {Lee} Y.~S.,  2014,
  \mn@doi [Science] {10.1126/science.1252633}, \href
  {http://adsabs.harvard.edu/abs/2014Sci...345..912A} {345, 912}

\bibitem[\protect\citeauthoryear{{Becker}, {Sargent}, {Rauch}  \&
  {Simcoe}}{{Becker} et~al.}{2006}]{Becker06}
{Becker} G.~D.,  {Sargent} W.~L.~W.,  {Rauch} M.,   {Simcoe} R.~A.,  2006,
  \mn@doi [\apj] {10.1086/500079}, \href
  {http://adsabs.harvard.edu/abs/2006ApJ...640...69B} {640, 69}

\bibitem[\protect\citeauthoryear{{Beers} \& {Christlieb}}{{Beers} \&
  {Christlieb}}{2005}]{Beers05}
{Beers} T.~C.,  {Christlieb} N.,  2005, \mn@doi [\araa]
  {10.1146/annurev.astro.42.053102.134057}, \href
  {http://adsabs.harvard.edu/abs/2005ARA%26A..43..531B} {43, 531}

\bibitem[\protect\citeauthoryear{{Beers}, {Preston}  \& {Shectman}}{{Beers}
  et~al.}{1985}]{Beers85}
{Beers} T.~C.,  {Preston} G.~W.,   {Shectman} S.~A.,  1985, \mn@doi [\aj]
  {10.1086/113917}, \href {http://adsabs.harvard.edu/abs/1985AJ.....90.2089B}
  {90, 2089}

\bibitem[\protect\citeauthoryear{{Beers}, {Preston}  \& {Shectman}}{{Beers}
  et~al.}{1992}]{Beers92}
{Beers} T.~C.,  {Preston} G.~W.,   {Shectman} S.~A.,  1992, \mn@doi [\aj]
  {10.1086/116207}, \href {http://adsabs.harvard.edu/abs/1992AJ....103.1987B}
  {103, 1987}

\bibitem[\protect\citeauthoryear{{Bensby}, {Feltzing}  \& {Oey}}{{Bensby}
  et~al.}{2014}]{Bensby14b}
{Bensby} T.,  {Feltzing} S.,   {Oey} M.~S.,  2014, \mn@doi [\aap]
  {10.1051/0004-6361/201322631}, \href
  {http://adsabs.harvard.edu/abs/2014A%26A...562A..71B} {562, A71}

\bibitem[\protect\citeauthoryear{{Bensby} et~al.,}{{Bensby}
  et~al.}{2017}]{Bensby17}
{Bensby} T.,  et~al., 2017, \mn@doi [\aap] {10.1051/0004-6361/201730560}, \href
  {http://adsabs.harvard.edu/abs/2017A%26A...605A..89B} {605, A89}

\bibitem[\protect\citeauthoryear{{Bland-Hawthorn} \&
  {Maloney}}{{Bland-Hawthorn} \& {Maloney}}{2001}]{Bland-Hawthorn01}
{Bland-Hawthorn} J.,  {Maloney} P.~R.,  2001, \mn@doi [\apjl] {10.1086/319654},
  \href {http://adsabs.harvard.edu/abs/2001ApJ...550L.231B} {550, L231}

\bibitem[\protect\citeauthoryear{{Bond}}{{Bond}}{1970}]{Bond70}
{Bond} H.~E.,  1970, \mn@doi [\apjs] {10.1086/190220}, \href
  {http://adsabs.harvard.edu/abs/1970ApJS...22..117B} {22, 117}

\bibitem[\protect\citeauthoryear{{Borthakur}, {Heckman}, {Leitherer}  \&
  {Overzier}}{{Borthakur} et~al.}{2014}]{Borthakur14}
{Borthakur} S.,  {Heckman} T.~M.,  {Leitherer} C.,   {Overzier} R.~A.,  2014,
  \mn@doi [Science] {10.1126/science.1254214}, \href
  {http://adsabs.harvard.edu/abs/2014Sci...346..216B} {346, 216}

\bibitem[\protect\citeauthoryear{{Bouwens}, {Illingworth}, {Oesch}, {Caruana},
  {Holwerda}, {Smit}  \& {Wilkins}}{{Bouwens} et~al.}{2015}]{Bouwens15}
{Bouwens} R.~J.,  {Illingworth} G.~D.,  {Oesch} P.~A.,  {Caruana} J.,
  {Holwerda} B.,  {Smit} R.,   {Wilkins} S.,  2015, \mn@doi [\apj]
  {10.1088/0004-637X/811/2/140}, \href
  {http://adsabs.harvard.edu/abs/2015ApJ...811..140B} {811, 140}

\bibitem[\protect\citeauthoryear{Bovill \& Ricotti}{Bovill \&
  Ricotti}{2011}]{Bovill11}
Bovill M.~S.,  Ricotti M.,  2011, The Astrophysical Journal, 741, 18

\bibitem[\protect\citeauthoryear{{Bovy}, {Rix}, {Schlafly}, {Nidever},
  {Holtzman}, {Shetrone}  \& {Beers}}{{Bovy} et~al.}{2016}]{Bovy16}
{Bovy} J.,  {Rix} H.-W.,  {Schlafly} E.~F.,  {Nidever} D.~L.,  {Holtzman}
  J.~A.,  {Shetrone} M.,   {Beers} T.~C.,  2016, \mn@doi [\apj]
  {10.3847/0004-637X/823/1/30}, \href
  {http://adsabs.harvard.edu/abs/2016ApJ...823...30B} {823, 30}

\bibitem[\protect\citeauthoryear{{Bridge} et~al.,}{{Bridge}
  et~al.}{2010}]{Bridge10}
{Bridge} C.~R.,  et~al., 2010, \mn@doi [\apj] {10.1088/0004-637X/720/1/465},
  \href {http://adsabs.harvard.edu/abs/2010ApJ...720..465B} {720, 465}

\bibitem[\protect\citeauthoryear{{Brisbin} \& {Harwit}}{{Brisbin} \&
  {Harwit}}{2012}]{Brisbin12}
{Brisbin} D.,  {Harwit} M.,  2012, \mn@doi [\apj]
  {10.1088/0004-637X/750/2/142}, \href
  {http://adsabs.harvard.edu/abs/2012ApJ...750..142B} {750, 142}

\bibitem[\protect\citeauthoryear{{Bromm} \& {Yoshida}}{{Bromm} \&
  {Yoshida}}{2011}]{Bromm11}
{Bromm} V.,  {Yoshida} N.,  2011, \mn@doi [\araa]
  {10.1146/annurev-astro-081710-102608}, \href
  {http://adsabs.harvard.edu/abs/2011ARA%26A..49..373B} {49, 373}

\bibitem[\protect\citeauthoryear{{Brook}, {Kawata}, {Scannapieco}, {Martel}  \&
  {Gibson}}{{Brook} et~al.}{2007}]{Brook07}
{Brook} C.~B.,  {Kawata} D.,  {Scannapieco} E.,  {Martel} H.,   {Gibson} B.~K.,
   2007, \mn@doi [\apj] {10.1086/511514}, \href
  {http://adsabs.harvard.edu/abs/2007ApJ...661...10B} {661, 10}

\bibitem[\protect\citeauthoryear{{Cai}, {Fan}, {Dave}, {Finlator}  \&
  {Oppenheimer}}{{Cai} et~al.}{2017}]{Cai17}
{Cai} Z.,  {Fan} X.,  {Dave} R.,  {Finlator} K.,   {Oppenheimer} B.,  2017,
  preprint, \href {http://adsabs.harvard.edu/abs/2017arXiv170910111C} {}
  (\mn@eprint {arXiv} {1709.10111})

\bibitem[\protect\citeauthoryear{{Carollo}, {Freeman}, {Beers}, {Placco},
  {Tumlinson}  \& {Martell}}{{Carollo} et~al.}{2014}]{Carollo14}
{Carollo} D.,  {Freeman} K.,  {Beers} T.~C.,  {Placco} V.~M.,  {Tumlinson} J.,
   {Martell} S.~L.,  2014, \mn@doi [\apj] {10.1088/0004-637X/788/2/180}, \href
  {http://adsabs.harvard.edu/abs/2014ApJ...788..180C} {788, 180}

\bibitem[\protect\citeauthoryear{{Chabrier}}{{Chabrier}}{2003}]{Chabrier03}
{Chabrier} G.,  2003, \mn@doi [\pasp] {10.1086/376392}, \href
  {http://adsabs.harvard.edu/abs/2003PASP..115..763C} {115, 763}

\bibitem[\protect\citeauthoryear{{Chan} \& {Heger}}{{Chan} \&
  {Heger}}{2016}]{Chan16}
{Chan} C.,  {Heger} A.,  2016, preprint, \href
  {http://adsabs.harvard.edu/abs/2016arXiv161006339C} {} (\mn@eprint {arXiv}
  {1610.06339})

\bibitem[\protect\citeauthoryear{{Chisholm}, {Orlitov{\'a}}, {Schaerer},
  {Verhamme}, {Worseck}, {Izotov}, {Thuan}  \& {Guseva}}{{Chisholm}
  et~al.}{2017}]{Chrisholm17}
{Chisholm} J.,  {Orlitov{\'a}} I.,  {Schaerer} D.,  {Verhamme} A.,  {Worseck}
  G.,  {Izotov} Y.~I.,  {Thuan} T.~X.,   {Guseva} N.~G.,  2017, \mn@doi [\aap]
  {10.1051/0004-6361/201730610}, \href
  {http://adsabs.harvard.edu/abs/2017A%26A...605A..67C} {605, A67}

\bibitem[\protect\citeauthoryear{{Christlieb}}{{Christlieb}}{2003}]{Christlieb03}
{Christlieb} N.,  2003, in {Schielicke} R.~E.,  ed.,  Reviews in Modern
  Astronomy Vol. 16, Reviews in Modern Astronomy. p.~191 (\mn@eprint {}
  {astro-ph/0308016}), \mn@doi{10.1002/9783527617647.ch8}

\bibitem[\protect\citeauthoryear{{Christlieb}}{{Christlieb}}{2016}]{Christlieb16}
{Christlieb} N.,  2016, \mn@doi [Astronomische Nachrichten]
  {10.1002/asna.201612401}, \href
  {http://adsabs.harvard.edu/abs/2016AN....337..931C} {337, 931}

\bibitem[\protect\citeauthoryear{{Christlieb} et~al.,}{{Christlieb}
  et~al.}{2002}]{Christlieb02}
{Christlieb} N.,  et~al., 2002, \mn@doi [\nat] {10.1038/nature01142}, \href
  {http://adsabs.harvard.edu/abs/2002Natur.419..904C} {419, 904}

\bibitem[\protect\citeauthoryear{{Cooke} \& {Madau}}{{Cooke} \&
  {Madau}}{2014}]{Cooke14}
{Cooke} R.~J.,  {Madau} P.,  2014, \mn@doi [\apj]
  {10.1088/0004-637X/791/2/116}, \href
  {http://adsabs.harvard.edu/abs/2014ApJ...791..116C} {791, 116}

\bibitem[\protect\citeauthoryear{{Cooke}, {Pettini}, {Steidel}, {Rudie}  \&
  {Nissen}}{{Cooke} et~al.}{2011}]{Cooke11}
{Cooke} R.,  {Pettini} M.,  {Steidel} C.~C.,  {Rudie} G.~C.,   {Nissen} P.~E.,
  2011, \mn@doi [\mnras] {10.1111/j.1365-2966.2011.19365.x}, \href
  {http://ukads.nottingham.ac.uk/abs/2011MNRAS.417.1534C} {417, 1534}

\bibitem[\protect\citeauthoryear{{Cooke}, {Pettini}  \& {Steidel}}{{Cooke}
  et~al.}{2017}]{Cooke17}
{Cooke} R.~J.,  {Pettini} M.,   {Steidel} C.~C.,  2017, \mn@doi [\mnras]
  {10.1093/mnras/stx037}, \href
  {http://adsabs.harvard.edu/abs/2017MNRAS.467..802C} {467, 802}

\bibitem[\protect\citeauthoryear{{Crain} et~al.,}{{Crain}
  et~al.}{2015}]{Crain15}
{Crain} R.~A.,  et~al., 2015, \mn@doi [\mnras] {10.1093/mnras/stv725}, \href
  {http://adsabs.harvard.edu/abs/2015MNRAS.450.1937C} {450, 1937}

\bibitem[\protect\citeauthoryear{{Dolag}, {Borgani}, {Murante}  \&
  {Springel}}{{Dolag} et~al.}{2009}]{Dolag09}
{Dolag} K.,  {Borgani} S.,  {Murante} G.,   {Springel} V.,  2009, \mn@doi
  [\mnras] {10.1111/j.1365-2966.2009.15034.x}, \href
  {http://adsabs.harvard.edu/abs/2009MNRAS.399..497D} {399, 497}

\bibitem[\protect\citeauthoryear{EagleTeam}{EagleTeam}{2017}]{Eagle17}
EagleTeam 2017, preprint, \href
  {http://adsabs.harvard.edu/abs/2017arXiv1706.09899} {} (\mn@eprint {arXiv}
  {1706.09899})

\bibitem[\protect\citeauthoryear{{El-Badry}, {Wetzel}, {Geha}, {Hopkins},
  {Kere{\v s}}, {Chan}  \& {Faucher-Gigu{\`e}re}}{{El-Badry}
  et~al.}{2016}]{ElBadry16}
{El-Badry} K.,  {Wetzel} A.,  {Geha} M.,  {Hopkins} P.~F.,  {Kere{\v s}} D.,
  {Chan} T.~K.,   {Faucher-Gigu{\`e}re} C.-A.,  2016, \mn@doi [\apj]
  {10.3847/0004-637X/820/2/131}, \href
  {http://adsabs.harvard.edu/abs/2016ApJ...820..131E} {820, 131}

\bibitem[\protect\citeauthoryear{{Frebel} \& {Norris}}{{Frebel} \&
  {Norris}}{2015}]{Frebel15}
{Frebel} A.,  {Norris} J.~E.,  2015, \mn@doi [\araa]
  {10.1146/annurev-astro-082214-122423}, \href
  {http://adsabs.harvard.edu/abs/2015ARA%26A..53..631F} {53, 631}

\bibitem[\protect\citeauthoryear{{Frebel}, {Collet}, {Eriksson}, {Christlieb}
  \& {Aoki}}{{Frebel} et~al.}{2008}]{Frebel08}
{Frebel} A.,  {Collet} R.,  {Eriksson} K.,  {Christlieb} N.,   {Aoki} W.,
  2008, \mn@doi [\apj] {10.1086/590327}, \href
  {http://adsabs.harvard.edu/abs/2008ApJ...684..588F} {684, 588}

\bibitem[\protect\citeauthoryear{{Frebel}, {Simon}  \& {Kirby}}{{Frebel}
  et~al.}{2014}]{Frebel14}
{Frebel} A.,  {Simon} J.~D.,   {Kirby} E.~N.,  2014, \mn@doi [\apj]
  {10.1088/0004-637X/786/1/74}, \href
  {http://adsabs.harvard.edu/abs/2014ApJ...786...74F} {786, 74}

\bibitem[\protect\citeauthoryear{{Furlong} et~al.,}{{Furlong}
  et~al.}{2017}]{Furlong17}
{Furlong} M.,  et~al., 2017, \mn@doi [\mnras] {10.1093/mnras/stw2740}, \href
  {http://adsabs.harvard.edu/abs/2017MNRAS.465..722F} {465, 722}

\bibitem[\protect\citeauthoryear{{Gaia Collaboration} et~al.,}{{Gaia
  Collaboration} et~al.}{2016}]{Gaia16}
{Gaia Collaboration} et~al., 2016, \mn@doi [\aap]
  {10.1051/0004-6361/201629272}, \href
  {http://adsabs.harvard.edu/abs/2016A%26A...595A...1G} {595, A1}

\bibitem[\protect\citeauthoryear{{Gardner} et~al.,}{{Gardner}
  et~al.}{2006}]{Gardner06}
{Gardner} J.~P.,  et~al., 2006, \mn@doi [\ssr] {10.1007/s11214-006-8315-7},
  \href {http://adsabs.harvard.edu/abs/2006SSRv..123..485G} {123, 485}

\bibitem[\protect\citeauthoryear{Gnedin \& Kravtsov}{Gnedin \&
  Kravtsov}{2006}]{Gnedin06}
Gnedin N.~Y.,  Kravtsov A.~V.,  2006, The Astrophysical Journal, 645, 1054

\bibitem[\protect\citeauthoryear{{Haardt} \& {Madau}}{{Haardt} \&
  {Madau}}{2001}]{Haardt01}
{Haardt} F.,  {Madau} P.,  2001, in {Neumann} D.~M.,  {Tran} J.~T.~V.,  eds,
  Clusters of Galaxies and the High Redshift Universe Observed in X-rays.
  (\mn@eprint {} {astro-ph/0106018})

\bibitem[\protect\citeauthoryear{{Haardt} \& {Madau}}{{Haardt} \&
  {Madau}}{2012}]{Haardt12}
{Haardt} F.,  {Madau} P.,  2012, \mn@doi [\apj] {10.1088/0004-637X/746/2/125},
  \href {http://adsabs.harvard.edu/abs/2012ApJ...746..125H} {746, 125}

\bibitem[\protect\citeauthoryear{{Haywood}, {Di Matteo}, {Snaith}  \&
  {Lehnert}}{{Haywood} et~al.}{2015}]{Haywood15}
{Haywood} M.,  {Di Matteo} P.,  {Snaith} O.,   {Lehnert} M.~D.,  2015, \mn@doi
  [\aap] {10.1051/0004-6361/201425459}, \href
  {http://adsabs.harvard.edu/abs/2015A%26A...579A...5H} {579, A5}

\bibitem[\protect\citeauthoryear{{Heckman}}{{Heckman}}{2001}]{Heckman01}
{Heckman} T.~M.,  2001, in {Hibbard} J.~E.,  {Rupen} M.,   {van Gorkom} J.~H.,
  eds,  Astronomical Society of the Pacific Conference Series Vol. 240, Gas and
  Galaxy Evolution. p.~345 (\mn@eprint {} {astro-ph/0009075})

\bibitem[\protect\citeauthoryear{{Helmi}, {Veljanoski}, {Breddels}, {Tian}  \&
  {Sales}}{{Helmi} et~al.}{2017}]{Helmi17}
{Helmi} A.,  {Veljanoski} J.,  {Breddels} M.~A.,  {Tian} H.,   {Sales} L.~V.,
  2017, \mn@doi [\aap] {10.1051/0004-6361/201629990}, \href
  {http://adsabs.harvard.edu/abs/2017A%26A...598A..58H} {598, A58}

\bibitem[\protect\citeauthoryear{{Ishigaki}, {Tominaga}, {Kobayashi}  \&
  {Nomoto}}{{Ishigaki} et~al.}{2014}]{Ishigaki14}
{Ishigaki} M.~N.,  {Tominaga} N.,  {Kobayashi} C.,   {Nomoto} K.,  2014,
  \mn@doi [\apjl] {10.1088/2041-8205/792/2/L32}, \href
  {http://ukads.nottingham.ac.uk/abs/2014ApJ...792L..32I} {792, L32}

\bibitem[\protect\citeauthoryear{{Izotov}, {Orlitov{\'a}}, {Schaerer}, {Thuan},
  {Verhamme}, {Guseva}  \& {Worseck}}{{Izotov} et~al.}{2016}]{Izotov16}
{Izotov} Y.~I.,  {Orlitov{\'a}} I.,  {Schaerer} D.,  {Thuan} T.~X.,  {Verhamme}
  A.,  {Guseva} N.~G.,   {Worseck} G.,  2016, \mn@doi [\nat]
  {10.1038/nature16456}, \href
  {http://adsabs.harvard.edu/abs/2016Natur.529..178I} {529, 178}

\bibitem[\protect\citeauthoryear{{Keller} et~al.,}{{Keller}
  et~al.}{2014}]{Keller14}
{Keller} S.~C.,  et~al., 2014, \mn@doi [\nat] {10.1038/nature12990}, \href
  {http://adsabs.harvard.edu/abs/2014Natur.506..463K} {506, 463}

\bibitem[\protect\citeauthoryear{{Kennicutt}}{{Kennicutt}}{1998}]{Kennicutt98}
{Kennicutt} Jr. R.~C.,  1998, \mn@doi [\araa] {10.1146/annurev.astro.36.1.189},
  \href {http://adsabs.harvard.edu/abs/1998ARA%26A..36..189K} {36, 189}

\bibitem[\protect\citeauthoryear{{Khaire}, {Srianand}, {Choudhury}  \&
  {Gaikwad}}{{Khaire} et~al.}{2015}]{Khaire15}
{Khaire} V.,  {Srianand} R.,  {Choudhury} T.~R.,   {Gaikwad} P.,  2015,
  preprint, \href {http://adsabs.harvard.edu/abs/2015arXiv151004700K} {}
  (\mn@eprint {arXiv} {1510.04700})

\bibitem[\protect\citeauthoryear{{Kimm} \& {Cen}}{{Kimm} \&
  {Cen}}{2014}]{Kimm14}
{Kimm} T.,  {Cen} R.,  2014, \mn@doi [\apj] {10.1088/0004-637X/788/2/121},
  \href {http://adsabs.harvard.edu/abs/2014ApJ...788..121K} {788, 121}

\bibitem[\protect\citeauthoryear{{Kunder} et~al.,}{{Kunder}
  et~al.}{2017}]{Kunder17}
{Kunder} A.,  et~al., 2017, \mn@doi [\aj] {10.3847/1538-3881/153/2/75}, \href
  {http://adsabs.harvard.edu/abs/2017AJ....153...75K} {153, 75}

\bibitem[\protect\citeauthoryear{{Lee} et~al.,}{{Lee} et~al.}{2013}]{Lee13}
{Lee} Y.~S.,  et~al., 2013, \mn@doi [\aj] {10.1088/0004-6256/146/5/132}, \href
  {http://adsabs.harvard.edu/abs/2013AJ....146..132L} {146, 132}

\bibitem[\protect\citeauthoryear{{Lee}, {Beers}, {Kim}, {Placco}, {Yoon},
  {Carollo}, {Masseron}  \& {Jung}}{{Lee} et~al.}{2017}]{Lee17}
{Lee} Y.~S.,  {Beers} T.~C.,  {Kim} Y.~K.,  {Placco} V.,  {Yoon} J.,  {Carollo}
  D.,  {Masseron} T.,   {Jung} J.,  2017, \mn@doi [\apj]
  {10.3847/1538-4357/836/1/91}, \href
  {http://adsabs.harvard.edu/abs/2017ApJ...836...91L} {836, 91}

\bibitem[\protect\citeauthoryear{{Leitherer}, {Hernandez}, {Lee}  \&
  {Oey}}{{Leitherer} et~al.}{2016}]{Leitherer16}
{Leitherer} C.,  {Hernandez} S.,  {Lee} J.~C.,   {Oey} M.~S.,  2016, \mn@doi
  [\apj] {10.3847/0004-637X/823/1/64}, \href
  {http://adsabs.harvard.edu/abs/2016ApJ...823...64L} {823, 64}

\bibitem[\protect\citeauthoryear{{Li}, {Zhao}, {Christlieb}, {Wang}, {Wang},
  {Zhang}, {Hou}  \& {Yuan}}{{Li} et~al.}{2015}]{Li15}
{Li} H.-N.,  {Zhao} G.,  {Christlieb} N.,  {Wang} L.,  {Wang} W.,  {Zhang} Y.,
  {Hou} Y.,   {Yuan} H.,  2015, \mn@doi [\apj] {10.1088/0004-637X/798/2/110},
  \href {http://adsabs.harvard.edu/abs/2015ApJ...798..110L} {798, 110}

\bibitem[\protect\citeauthoryear{{Madau} \& {Haardt}}{{Madau} \&
  {Haardt}}{2015}]{Haardt15}
{Madau} P.,  {Haardt} F.,  2015, \mn@doi [\apjl] {10.1088/2041-8205/813/1/L8},
  \href {http://adsabs.harvard.edu/abs/2015ApJ...813L...8M} {813, L8}

\bibitem[\protect\citeauthoryear{Madau, Kuhlen, Diemand, Moore, Zemp, Potter
  \& Stadel}{Madau et~al.}{2008}]{Madau08}
Madau P.,  Kuhlen M.,  Diemand J.,  Moore B.,  Zemp M.,  Potter D.,   Stadel
  J.,  2008, The Astrophysical Journal Letters, 689, L41

\bibitem[\protect\citeauthoryear{{Marchi} et~al.,}{{Marchi}
  et~al.}{2017}]{Marchi17}
{Marchi} F.,  et~al., 2017, preprint, \href
  {http://adsabs.harvard.edu/abs/2017arXiv171010184M} {} (\mn@eprint {arXiv}
  {1710.10184})

\bibitem[\protect\citeauthoryear{{Martig} et~al.,}{{Martig}
  et~al.}{2016}]{Martig16}
{Martig} M.,  et~al., 2016, \mn@doi [\mnras] {10.1093/mnras/stv2830}, \href
  {http://adsabs.harvard.edu/abs/2016MNRAS.456.3655M} {456, 3655}

\bibitem[\protect\citeauthoryear{{McAlpine} et~al.,}{{McAlpine}
  et~al.}{2016}]{McAlpine16}
{McAlpine} S.,  et~al., 2016, \mn@doi [Astronomy and Computing]
  {10.1016/j.ascom.2016.02.004}, \href
  {http://adsabs.harvard.edu/abs/2016A%26C....15...72M} {15, 72}

\bibitem[\protect\citeauthoryear{{Mostardi}, {Shapley}, {Steidel}, {Trainor},
  {Reddy}  \& {Siana}}{{Mostardi} et~al.}{2015}]{Mostardi15}
{Mostardi} R.~E.,  {Shapley} A.~E.,  {Steidel} C.~C.,  {Trainor} R.~F.,
  {Reddy} N.~A.,   {Siana} B.,  2015, \mn@doi [\apj]
  {10.1088/0004-637X/810/2/107}, \href
  {http://adsabs.harvard.edu/abs/2015ApJ...810..107M} {810, 107}

\bibitem[\protect\citeauthoryear{{Navarro}, {Frenk}  \& {White}}{{Navarro}
  et~al.}{1996}]{Navarro96}
{Navarro} J.~F.,  {Frenk} C.~S.,   {White} S.~D.~M.,  1996, \mn@doi [\apj]
  {10.1086/177173}, \href {http://adsabs.harvard.edu/abs/1996ApJ...462..563N}
  {462, 563}

\bibitem[\protect\citeauthoryear{{Navarro} et~al.,}{{Navarro}
  et~al.}{2017}]{Navarro17}
{Navarro} J.~F.,  et~al., 2017, preprint, \href
  {http://adsabs.harvard.edu/abs/2017arXiv170901040N} {} (\mn@eprint {arXiv}
  {1709.01040})

\bibitem[\protect\citeauthoryear{{Nomoto}, {Kobayashi}  \& {Tominaga}}{{Nomoto}
  et~al.}{2013}]{Nomoto13}
{Nomoto} K.,  {Kobayashi} C.,   {Tominaga} N.,  2013, \mn@doi [\araa]
  {10.1146/annurev-astro-082812-140956}, \href
  {http://adsabs.harvard.edu/abs/2013ARA%26A..51..457N} {51, 457}

\bibitem[\protect\citeauthoryear{{Norris} et~al.,}{{Norris}
  et~al.}{2013}]{Norris13a}
{Norris} J.~E.,  et~al., 2013, \mn@doi [\apj] {10.1088/0004-637X/762/1/28},
  \href {http://adsabs.harvard.edu/abs/2013ApJ...762...28N} {762, 28}

\bibitem[\protect\citeauthoryear{{Parsa}, {Dunlop}  \& {McLure}}{{Parsa}
  et~al.}{2017}]{Parsa17}
{Parsa} S.,  {Dunlop} J.~S.,   {McLure} R.~J.,  2017, preprint, \href
  {http://adsabs.harvard.edu/abs/2017arXiv170407750P} {} (\mn@eprint {arXiv}
  {1704.07750})

\bibitem[\protect\citeauthoryear{{Pian} et~al.,}{{Pian} et~al.}{2017}]{Pian17}
{Pian} E.,  et~al., 2017, preprint, \href
  {http://adsabs.harvard.edu/abs/2017arXiv171005858P} {} (\mn@eprint {arXiv}
  {1710.05858})

\bibitem[\protect\citeauthoryear{{Pilkington} et~al.,}{{Pilkington}
  et~al.}{2012}]{Pilkington12}
{Pilkington} K.,  et~al., 2012, \mn@doi [\mnras]
  {10.1111/j.1365-2966.2012.21353.x}, \href
  {http://adsabs.harvard.edu/abs/2012MNRAS.425..969P} {425, 969}

\bibitem[\protect\citeauthoryear{{Planck Collaboration} et~al.,}{{Planck
  Collaboration} et~al.}{2015}]{Planck15}
{Planck Collaboration} et~al., 2015, preprint, \href
  {http://adsabs.harvard.edu/abs/2015arXiv150201589P} {} (\mn@eprint {arXiv}
  {1502.01589})

\bibitem[\protect\citeauthoryear{{Robertson}, {Ellis}, {Furlanetto}  \&
  {Dunlop}}{{Robertson} et~al.}{2015}]{Robertson15}
{Robertson} B.~E.,  {Ellis} R.~S.,  {Furlanetto} S.~R.,   {Dunlop} J.~S.,
  2015, \mn@doi [\apjl] {10.1088/2041-8205/802/2/L19}, \href
  {http://adsabs.harvard.edu/abs/2015ApJ...802L..19R} {802, L19}

\bibitem[\protect\citeauthoryear{{Salvadori}, {Ferrara}, {Schneider},
  {Scannapieco}  \& {Kawata}}{{Salvadori} et~al.}{2010}]{Salvadori10}
{Salvadori} S.,  {Ferrara} A.,  {Schneider} R.,  {Scannapieco} E.,   {Kawata}
  D.,  2010, \mn@doi [\mnras] {10.1111/j.1745-3933.2009.00772.x}, \href
  {http://adsabs.harvard.edu/abs/2010MNRAS.401L...5S} {401, L5}

\bibitem[\protect\citeauthoryear{{Sarmento}, {Scannapieco}  \&
  {Pan}}{{Sarmento} et~al.}{2017}]{Sarmento17}
{Sarmento} R.,  {Scannapieco} E.,   {Pan} L.,  2017, \mn@doi [\apj]
  {10.3847/1538-4357/834/1/23}, \href
  {http://adsabs.harvard.edu/abs/2017ApJ...834...23S} {834, 23}

\bibitem[\protect\citeauthoryear{{Sawala} et~al.,}{{Sawala}
  et~al.}{2016}]{Sawala16}
{Sawala} T.,  et~al., 2016, \mn@doi [\mnras] {10.1093/mnras/stw145}, \href
  {http://adsabs.harvard.edu/abs/2016MNRAS.457.1931S} {457, 1931}

\bibitem[\protect\citeauthoryear{{Schaye} \& {Dalla Vecchia}}{{Schaye} \&
  {Dalla Vecchia}}{2008}]{Schaye08}
{Schaye} J.,  {Dalla Vecchia} C.,  2008, \mn@doi [\mnras]
  {10.1111/j.1365-2966.2007.12639.x}, \href
  {http://adsabs.harvard.edu/abs/2008MNRAS.383.1210S} {383, 1210}

\bibitem[\protect\citeauthoryear{{Schaye} et~al.,}{{Schaye}
  et~al.}{2015}]{Schaye15}
{Schaye} J.,  et~al., 2015, \mn@doi [\mnras] {10.1093/mnras/stu2058}, \href
  {http://adsabs.harvard.edu/abs/2015MNRAS.446..521S} {446, 521}

\bibitem[\protect\citeauthoryear{{Schechter}}{{Schechter}}{1976}]{Schechter76}
{Schechter} P.,  1976, \mn@doi [\apj] {10.1086/154079}, \href
  {http://adsabs.harvard.edu/abs/1976ApJ...203..297S} {203, 297}

\bibitem[\protect\citeauthoryear{{Sellwood} \& {Binney}}{{Sellwood} \&
  {Binney}}{2002}]{Sellwood02}
{Sellwood} J.~A.,  {Binney} J.~J.,  2002, \mn@doi [\mnras]
  {10.1046/j.1365-8711.2002.05806.x}, \href
  {http://adsabs.harvard.edu/abs/2002MNRAS.336..785S} {336, 785}

\bibitem[\protect\citeauthoryear{{Sharma}, {Theuns}, {Frenk}  \&
  {Cooke}}{{Sharma} et~al.}{2016a}]{Sharma16b}
{Sharma} M.,  {Theuns} T.,  {Frenk} C.,   {Cooke} R.,  2016a, preprint, \href
  {http://adsabs.harvard.edu/abs/2016arXiv161103868S} {} (\mn@eprint {arXiv}
  {1611.03868})

\bibitem[\protect\citeauthoryear{{Sharma}, {Theuns}, {Frenk}, {Bower}, {Crain},
  {Schaller}  \& {Schaye}}{{Sharma} et~al.}{2016b}]{Sharma16}
{Sharma} M.,  {Theuns} T.,  {Frenk} C.,  {Bower} R.,  {Crain} R.,  {Schaller}
  M.,   {Schaye} J.,  2016b, \mn@doi [\mnras] {10.1093/mnrasl/slw021}, \href
  {http://adsabs.harvard.edu/abs/2016MNRAS.458L..94S} {458, L94}

\bibitem[\protect\citeauthoryear{{Sharma}, {Theuns}, {Frenk}, {Bower}, {Crain},
  {Schaller}  \& {Schaye}}{{Sharma} et~al.}{2017}]{Sharma17}
{Sharma} M.,  {Theuns} T.,  {Frenk} C.,  {Bower} R.~G.,  {Crain} R.~A.,
  {Schaller} M.,   {Schaye} J.,  2017, \mn@doi [\mnras] {10.1093/mnras/stx578},
  \href {http://adsabs.harvard.edu/abs/2017MNRAS.468.2176S} {468, 2176}

\bibitem[\protect\citeauthoryear{{Shibuya}, {Ouchi}  \& {Harikane}}{{Shibuya}
  et~al.}{2015}]{Shibuya15}
{Shibuya} T.,  {Ouchi} M.,   {Harikane} Y.,  2015, \mn@doi [\apjs]
  {10.1088/0067-0049/219/2/15}, \href
  {http://adsabs.harvard.edu/abs/2015ApJS..219...15S} {219, 15}

\bibitem[\protect\citeauthoryear{{Soderblom}}{{Soderblom}}{2010}]{Soderblom10}
{Soderblom} D.~R.,  2010, \mn@doi [\araa]
  {10.1146/annurev-astro-081309-130806}, \href
  {http://adsabs.harvard.edu/abs/2010ARA%26A..48..581S} {48, 581}

\bibitem[\protect\citeauthoryear{{Spite} et~al.,}{{Spite}
  et~al.}{2011}]{Caffau11b}
{Spite} M.,  et~al., 2011, \mn@doi [\aap] {10.1051/0004-6361/201015926}, \href
  {http://adsabs.harvard.edu/abs/2011A%26A...528A...9S} {528, A9}

\bibitem[\protect\citeauthoryear{{Springel}}{{Springel}}{2005}]{Springel05}
{Springel} V.,  2005, \mn@doi [\mnras] {10.1111/j.1365-2966.2005.09655.x},
  \href {http://adsabs.harvard.edu/abs/2005MNRAS.364.1105S} {364, 1105}

\bibitem[\protect\citeauthoryear{{Springel}, {White}, {Tormen}  \&
  {Kauffmann}}{{Springel} et~al.}{2001}]{Springel01}
{Springel} V.,  {White} S.~D.~M.,  {Tormen} G.,   {Kauffmann} G.,  2001,
  \mn@doi [\mnras] {10.1046/j.1365-8711.2001.04912.x}, \href
  {http://adsabs.harvard.edu/abs/2001MNRAS.328..726S} {328, 726}

\bibitem[\protect\citeauthoryear{Starkenburg, Oman, Navarro, Crain, Fattahi,
  Frenk, Sawala  \& Schaye}{Starkenburg et~al.}{2017}]{Starkenburg17}
Starkenburg E.,  Oman K.~A.,  Navarro J.~F.,  Crain R.~A.,  Fattahi A.,  Frenk
  C.~S.,  Sawala T.,   Schaye J.,  2017, Monthly Notices of the Royal
  Astronomical Society, 465, 2212

\bibitem[\protect\citeauthoryear{{Suda}, {Yamada}, {Katsuta}, {Komiya},
  {Ishizuka}, {Aoki}  \& {Fujimoto}}{{Suda} et~al.}{2011}]{Suda11}
{Suda} T.,  {Yamada} S.,  {Katsuta} Y.,  {Komiya} Y.,  {Ishizuka} C.,  {Aoki}
  W.,   {Fujimoto} M.~Y.,  2011, \mn@doi [\mnras]
  {10.1111/j.1365-2966.2011.17943.x}, \href
  {http://ads.nao.ac.jp/abs/2011MNRAS.412..843S} {412, 843}

\bibitem[\protect\citeauthoryear{{Thielemann}, {Eichler}, {Panov}  \&
  {Wehmeyer}}{{Thielemann} et~al.}{2017}]{Thieleman17}
{Thielemann} K. F.,  {Eichler} M.,  {Panov} I.~V.,   {Wehmeyer} B.,  2017,
  \mn@doi [Annual Review of Nuclear and Particle Science]
  {10.1146/annurev-nucl-101916-123246}, \href
  {http://adsabs.harvard.edu/abs/2017ARNPS..6701916T} {67, annurev}

\bibitem[\protect\citeauthoryear{{Trebitsch}, {Blaizot}, {Rosdahl}, {Devriendt}
   \& {Slyz}}{{Trebitsch} et~al.}{2017}]{Trebitsch17}
{Trebitsch} M.,  {Blaizot} J.,  {Rosdahl} J.,  {Devriendt} J.,   {Slyz} A.,
  2017, \mn@doi [\mnras] {10.1093/mnras/stx1060}, \href
  {http://adsabs.harvard.edu/abs/2017MNRAS.470..224T} {470, 224}

\bibitem[\protect\citeauthoryear{{Trenti}, {Stiavelli}, {Bouwens}, {Oesch},
  {Shull}, {Illingworth}, {Bradley}  \& {Carollo}}{{Trenti}
  et~al.}{2010}]{Trenti10}
{Trenti} M.,  {Stiavelli} M.,  {Bouwens} R.~J.,  {Oesch} P.,  {Shull} J.~M.,
  {Illingworth} G.~D.,  {Bradley} L.~D.,   {Carollo} C.~M.,  2010, \mn@doi
  [\apjl] {10.1088/2041-8205/714/2/L202}, \href
  {http://adsabs.harvard.edu/abs/2010ApJ...714L.202T} {714, L202}

\bibitem[\protect\citeauthoryear{{Tumlinson}}{{Tumlinson}}{2010}]{Tumlinson10}
{Tumlinson} J.,  2010, \mn@doi [\apj] {10.1088/0004-637X/708/2/1398}, \href
  {http://adsabs.harvard.edu/abs/2010ApJ...708.1398T} {708, 1398}

\bibitem[\protect\citeauthoryear{{Wang}, {Han}, {Cooper}, {Cole}, {Frenk}  \&
  {Lowing}}{{Wang} et~al.}{2015}]{Wang15}
{Wang} W.,  {Han} J.,  {Cooper} A.~P.,  {Cole} S.,  {Frenk} C.,   {Lowing} B.,
  2015, \mn@doi [\mnras] {10.1093/mnras/stv1647}, \href
  {http://adsabs.harvard.edu/abs/2015MNRAS.453..377W} {453, 377}

\bibitem[\protect\citeauthoryear{{Webster}, {Frebel}  \&
  {Bland-Hawthorn}}{{Webster} et~al.}{2016}]{Webster16}
{Webster} D.,  {Frebel} A.,   {Bland-Hawthorn} J.,  2016, \mn@doi [\apj]
  {10.3847/0004-637X/818/1/80}, \href
  {http://adsabs.harvard.edu/abs/2016ApJ...818...80W} {818, 80}

\bibitem[\protect\citeauthoryear{{White} \& {Springel}}{{White} \&
  {Springel}}{2000}]{White00}
{White} S.~D.~M.,  {Springel} V.,  2000, in {Weiss} A.,  {Abel} T.~G.,   {Hill}
  V.,  eds, The First Stars. p.~327 (\mn@eprint {} {astro-ph/9911378}),
  \mn@doi{10.1007/10719504_62}

\bibitem[\protect\citeauthoryear{{Wiersma}, {Schaye}  \& {Smith}}{{Wiersma}
  et~al.}{2009a}]{Wiersma09a}
{Wiersma} R.~P.~C.,  {Schaye} J.,   {Smith} B.~D.,  2009a, \mn@doi [\mnras]
  {10.1111/j.1365-2966.2008.14191.x}, \href
  {http://adsabs.harvard.edu/abs/2009MNRAS.393...99W} {393, 99}

\bibitem[\protect\citeauthoryear{{Wiersma}, {Schaye}, {Theuns}, {Dalla Vecchia}
   \& {Tornatore}}{{Wiersma} et~al.}{2009b}]{Wiersma09b}
{Wiersma} R.~P.~C.,  {Schaye} J.,  {Theuns} T.,  {Dalla Vecchia} C.,
  {Tornatore} L.,  2009b, \mn@doi [\mnras] {10.1111/j.1365-2966.2009.15331.x},
  \href {http://adsabs.harvard.edu/abs/2009MNRAS.399..574W} {399, 574}

\bibitem[\protect\citeauthoryear{{Williamson}, {Martel}  \&
  {Kawata}}{{Williamson} et~al.}{2016}]{Williamson16}
{Williamson} D.,  {Martel} H.,   {Kawata} D.,  2016, \mn@doi [\apj]
  {10.3847/0004-637X/822/2/91}, \href
  {http://adsabs.harvard.edu/abs/2016ApJ...822...91W} {822, 91}

\bibitem[\protect\citeauthoryear{{Wise} \& {Cen}}{{Wise} \&
  {Cen}}{2009}]{Wise09}
{Wise} J.~H.,  {Cen} R.,  2009, \mn@doi [\apj] {10.1088/0004-637X/693/1/984},
  \href {http://adsabs.harvard.edu/abs/2009ApJ...693..984W} {693, 984}

\bibitem[\protect\citeauthoryear{{Yoon} et~al.,}{{Yoon} et~al.}{2016}]{Yoon16}
{Yoon} J.,  et~al., 2016, ArXiv e-prints: 1607.06336, \href
  {http://adsabs.harvard.edu/abs/2016arXiv160706336Y} {}

\bibitem[\protect\citeauthoryear{{Yoon}, {Beers}, {Kim}, {Placco}, {Yoon},
  {Carollo}, {Masseron}  \& {Sharma}}{{Yoon} et~al.}{2018}]{Yoon18}
{Yoon} J.,  {Beers} T.~C.,  {Kim} Y.~K.,  {Placco} V.,  {Yoon} J.,  {Carollo}
  D.,  {Masseron} T.,   {Sharma} M.,  2018, \apj, 836, 91

\bibitem[\protect\citeauthoryear{{Zackrisson} et~al.,}{{Zackrisson}
  et~al.}{2017}]{Zackrisson17}
{Zackrisson} E.,  et~al., 2017, \mn@doi [\apj] {10.3847/1538-4357/836/1/78},
  \href {http://adsabs.harvard.edu/abs/2017ApJ...836...78Z} {836, 78}

\makeatother
\end{thebibliography}

\section*{Appendix: supernovae yields and the distribution of [$\alpha$/F\MakeLowercase{e}]}
\label{sect:Appendix1}
\begin{figure}
 \centering
 \includegraphics[width=\linewidth]{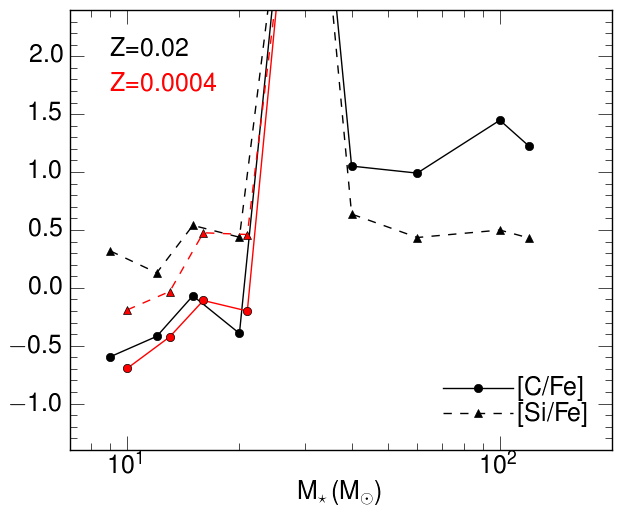}
\caption{The net abundance ratios of [C/Fe] (solid lines) and [Si/Fe] (dashed lines) in the ejecta of massive stars and their supernova descendants, as a function of progenitor stellar mass and shown for two values of the progenitor metallicity. These are the yields used in \eagle\ and are described in detail by \protect\cite{Wiersma09b}.  [C/Fe] yields are sub-solar for low-mass SNe, even at low $Z$. At higher progenitor masses, [C/Fe] becomes super-solar.  In contrast, [Si/Fe] is always at least close to solar, and often super-solar. At low $Z$, Fe does not escape the core during the SN explosion, and the ejecta are highly enriched in both C and Si compared to Fe, in the limiting case ejecta do not contain Fe but only lighter $\alpha$ elements.}
 \label{fig_Y}
\end{figure}

\begin{figure*}
 \centering
 \includegraphics[width=0.8\linewidth]{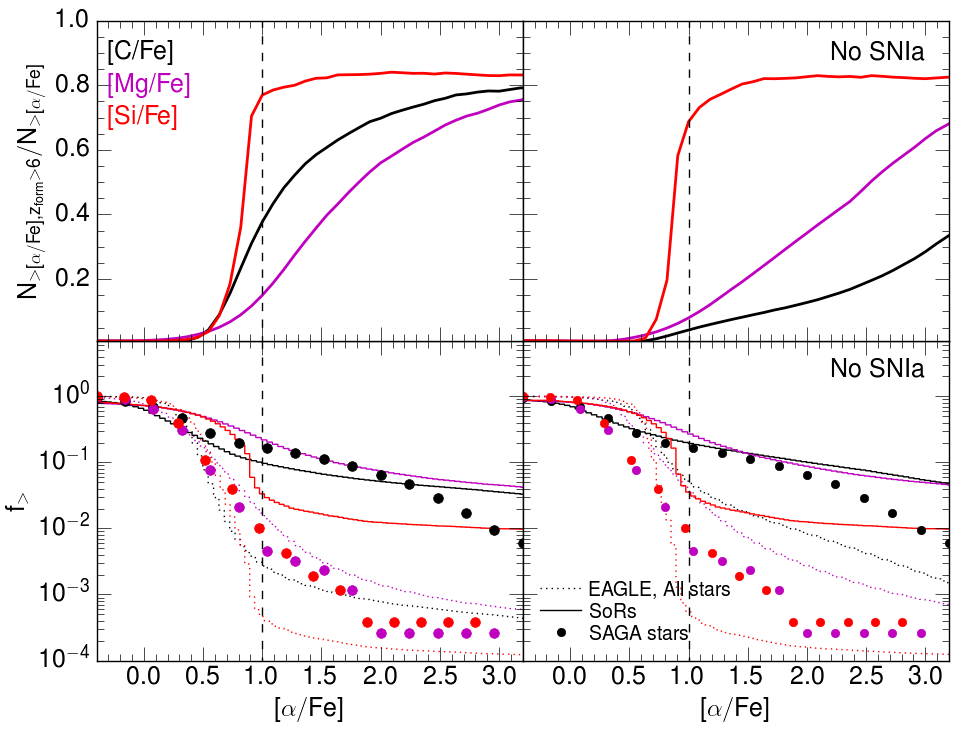}
\caption{
Cumulative distribution of the abundance ratios for different $\alpha$-elements in \eagle\ MW stars. {\em Bottom left panel:} Distribution of [C/Fe] ({\em black}), [Mg/Fe] ({\em purple}) and [Si/Fe] ({\em red}) for all stars (dashed lines) and for SoRs ({\em i.e.} formed before redshift $z=6$, full lines). Observed stars taken from the {\sc saga} database \citep{Suda11} are plotted as symbols, {\em black circles} show [C/Fe], {\em purple circles} show [Mg/Fe], and {\em red circles} show [Si/Fe]. {\em Top left panel:} Fraction of \eagle\ stars with a given value of [$\alpha$/Fe] that are SoRs, using the same colour scheme. Right panels show the same curves for \eagle\, but after removing the contribution from type~Ia SNe. The {\sc saga} data is repeated to guide the eye.}
 \label{fig_alphaFe}
\end{figure*}

As discussed in the main text, the oldest stars were mostly enriched by low metallicity type~II SNe. The yields for massive stars and their associated supernovae, as used in \eagle, are described in detail by \cite{Wiersma09b} and are shown in Fig.~\ref{fig_Y} for two different values of the progenitor metallicity $Z$ as a function of initial stellar mass. The net ejecta of these massive stars is typically super-solar in both [C/Fe] and [Si/Fe] (as well as other $\alpha$-elements), in particular for massive progenitors stars at low $Z$.

Enrichment in \eagle\ is time resolved, meaning that a recently formed star particle will enrich its gas neighbours with the yields from its most massive progenitor stars first, before enriching it with ejecta from lower mass SNe. It is thus possible for a gas particle to be enriched with produce that originates {\em only} from, say, very massive $Z\sim 0$ stars.  In the onion model for the SN precursor, $\alpha$ elements with higher atomic number A lie deeper in the stellar interior. These deeper layers may not be able to escape fall back after the star explodes, which is thought to happen at low initial $Z$ when the precursor massive star is more strongly bound. This model therefore predicts that low-$Z$ SNe have yields that are very high in [C/Fe] (potentially only producing C and no Fe at all), but less so in [Mg/Fe] and even less in in [Si/Fe] (because eventually neither the Si, nor the Fe core, is blasted into space). The enrichment implementation in \eagle\ should be able to capture this, with the caveat that injection of SN energy is currently {\em not} time resolved: all energy associated with SNe is injected 30~Myears after the star particle formed. As a consequence it is possible that most gas particles are typically enriched by the whole initial mass function worth of massive stars, because the enriched gas particle remains near the star particle.

The resulting enrichment pattern in \eagle\ is analysed in more detail in Fig.~\ref{fig_alphaFe}:
as expected, a relatively significant fraction of stars that form early (SoRs, formed before $z=6$) is highly over abundant in [C/Fe], [Mg/Fe] and [Si/Fe] (full lines in bottom left panel) as compared to all stars in $z=0$ MW galaxies (corresponding dotted lines). Comparing the trends for the different $\alpha$ elements in SoRs in more detail, we notice that C and Mg track each other much better than that they track Si, consistent with Si having much higher atomic number: for a range of progenitor mass and initial metallicity, the lighter $\alpha$ elements C and Mg are ejecta from the SN, when Si, and to a much larger extent Fe, are not. This is the origin of the very high [C/Fe] and [Mg/Fe] stars, while there are far fewer high [Si/Fe] stars. Also striking is the sharp upturn in the [Si/Fe] curve around [Si/Fe]=1, below which both Si and Fe are ejected yielding near solar [Si/Fe]. The transition from supersolar to near solar abundances in [C/Fe] and [Mg/Fe] is much more gentle. The different behaviour of Si compared to Mg and C is also clear from the top left panel of Fig.~\ref{fig_alphaFe}: very high overabundances in [Si/Fe] are very much restricted to the very earliest episodes of star formation, with nearly 80~per cent of stars with [Si/Fe]$>1$ formed before $z=6$. In contrast, only 40 and 15~per cent of stars formed before $z=6$ have  [Mg/Fe]$>1$ and [C/Fe]$>1$, respectively.

Interestingly, although the \eagle\ $\alpha$-element patterns exhibit the behaviour expected from the onion model of SN progenitors, if anything the trend is more pronounced in the {\sc saga} data \citep{Suda11}. In the observations, we do indeed find stars very highly enriched in [C/Fe], as we did in \eagle\, but the fraction of stars very highly obver abundant in [Si/Fe] is much lower, consistent with Si enrichment tracking Fe better than C enrichment does. Similarly many fewer {\sc saga} stars have extremely high [Mg/Fe] compared to \eagle. 

In addition to biases and lack of completeness in the data, there are several possible reasons for the discrepancy, including 	({\em i}) exaggerated lack of pollution of early stars with Fe from type~Ia in \eagle, ({\em ii}) lack of mixing of gas enriched by type~II SNe with different progenitor masses, ({\em iii}) issues with the yields of massive stars used, or ({\em iv}) differences in the initial stellar initial mass function. The right panels in Fig.~\ref{fig_alphaFe} show that the impact of type~Ia SNe is not large in \eagle, although of course we could have {\em underestimated} the amount of Fe produced by these stars - and how well that Fe is mixed with gas enriched by type~II SNe. Such effects due to lack of mixing are difficult to quantify, but it is striking how well the fraction of stars with a given [C/Fe] in SoRs tracks that measured in {\sc saga}. In the data we see a relatively pronounced upturn in the curves of [Mg/Fe] and [Si/Fe], pointing to a substantial increase in the fraction of SoRs with abundances of [Mg/Fe]$\lesssim 0.5$ and [Si/Fe]$\lesssim 0.5$. 
For [Si/Fe], this upturn is reminiscent of the much sharper upturn which occurs at [Si/Fe]$\sim 0.9$, possibly a hint that low-$Z$ SNe have too high Si yields in the simulation. In the simulation, this is even more the case for Mg.

% The ejecta of the low metallicity type-II supernovae is iron poor and therefore the yields of such SNe are typically rich in [$\alpha$/Fe]. Indeed, as we see in Fig.~\ref{fig_Y}, the low-Z SNe with low progenitor mass lower than  20 M$_\odot$ that have been implemented in \eagle\, have supersolar [C/Fe] and [Si/Fe]. Another, interesting fact  that follows from the onion model of the SNe explosion is that the yields of the $\alpha$-elements should decrease with increasing atomic number, from C to Si. We see this trend in the metal poor stars compiled in {\sc saga} database \citep{Suda11}, but in \eagle the this trend is not prominent. 

% There is a good agreement when comparing the distribution of [C/Fe] in \eagle\  with that from {\sc saga}, however  we find an excess of SoRs with high [Mg/Fe] and [Si/Fe]. Is this because of the SN type-Ia yields? However we find a negligible improvement even when we remove the contribution from SN type-Ia (see the right panels in Fig.~\ref{fig_alphaFe}). Instead, the answer lies in the unusually higher values of [Mg/Fe] and [Si/Fe] yields that we have used in the \eagle\ simulation. As we see in Fig.~\ref{fig_Y}, the implemented  yields in [Si/Fe] are a little bit higher.

\end{document}